\def\year{2022}\relax
\documentclass[letterpaper]{article} 
\usepackage{aaai22}  
\usepackage{times}  
\usepackage{helvet}  
\usepackage{courier}  
\usepackage[hyphens]{url}  
\usepackage{graphicx} 
\usepackage{natbib}  
\usepackage{caption} 
\DeclareCaptionStyle{ruled}{labelfont=normalfont,labelsep=colon,strut=off} 
\usepackage[ruled,vlined,linesnumbered]{algorithm2e}
\usepackage{graphicx}
\usepackage{caption}
\usepackage{subcaption}
\usepackage{amsfonts}
\usepackage{amsmath}
\usepackage{booktabs}
\usepackage{siunitx}
\usepackage{multirow}
\usepackage{xspace}
\usepackage{easy-todo}
\usepackage{tikz}
\usepackage{comment}
\usepackage{tabularx}
\usepackage{mathtools, nccmath}
\usepackage[shortlabels]{enumitem}
\usepackage{listings}
\usepackage{xpatch}
\usepackage[switch]{lineno}
\usepackage{makecell}

\usepackage[page]{appendix} 

\newcommand{\name}{{\sc Electra}\xspace}
\newcommand{\naru}{{\sc Naru}\xspace}
\newcommand*\circled[1]{\tikz[baseline=(char.base)]{\node[shape=circle,fill,inner sep=1pt] (char) {\textcolor{white}{#1}}}}

\iftrue
\newcommand{\sm}[1]{{\color{blue}{\small{\bf [Subrata: #1]}}}}
\else
\newcommand{\sm}[1]{}
\fi

\title{Conditional Generative Model based Predicate-Aware Query Approximation}
\author{
Nikhil Sheoran\textsuperscript{\rm 1}\thanks{Work done while at Adobe Research}, 
Subrata Mitra\textsuperscript{\rm 2}\thanks{Corresponding Author}, 
Vibhor Porwal\textsuperscript{\rm 2}, 
Siddharth Ghetia\textsuperscript{\rm 3*},\\
Jatin Varshney\textsuperscript{\rm 3*}, 
Tung Mai\textsuperscript{\rm 2}, 
Anup Rao\textsuperscript{\rm 2}, 
Vikas Maddukuri\textsuperscript{\rm 3*}
}
\affiliations{
\textsuperscript{\rm 1} University of Illinois at Urbana-Champaign
\hspace{1em}
\textsuperscript{\rm 2} Adobe Research
\hspace{1em}
\textsuperscript{\rm 3} Indian Institute of Technology, Roorkee
}
\begin{document}
\maketitle




\begin{abstract}
The goal of Approximate Query Processing (AQP) is to provide very fast but ``accurate enough'' results for costly aggregate queries thereby improving user experience in interactive exploration of large datasets.
Recently proposed Machine-Learning-based AQP techniques can provide very low latency as query execution only involves model inference as compared to traditional query processing on database clusters. However, with increase in the number of filtering predicates (\texttt{WHERE} clauses), the approximation error significantly increases for these methods. Analysts often use queries with a large number of predicates for insights discovery. Thus, maintaining low approximation error is important to prevent analysts from drawing misleading conclusions. In this paper, we propose \name\footnote{For more details: \url{https://nikhil96sher.github.io/electra}}, a predicate-aware AQP system that can answer analytics-style queries with a large number of predicates with much smaller approximation errors. \name uses a conditional generative model that learns the conditional distribution of the data and at run-time generates a small ($\approx$ 1000 rows) but representative sample, on which the query is executed to compute the approximate result. Our evaluations with four different baselines on three real-world datasets show that \name provides lower AQP error for large number of predicates compared to baselines.
\end{abstract}
\maketitle
\section{Introduction}
Interactive exploration and visualization tools, such as Tableau, Microsoft Power BI, Qlik, Polaris \cite{polaris}, and Vizdom \cite{vizdom} have gained popularity amongst data-analysts.
One of the desirable properties of these tools is that the speed of interaction with the data, i.e., the queries and the corresponding visualizations must complete at \textit{``human speed''}~
\cite{2016_IDEA} or at \textit{``rates resonant with the pace of human thought''}~\cite{latency-infoviz}.

To reduce the latency of such interactions, Approximate Query Processing (AQP) techniques that can provide fast but ``accurate enough'' results for queries with aggregates (AVG, SUM, COUNT) on numerical attributes on a large dataset, have recently gained popularity~\cite{ deepdb, vae, dbest}.
Prior AQP systems mainly relied on sampling-based methods. For example, BlinkDB~\cite{blinkdb} and recently proposed VerdictDB~\cite{verdictdb}, first create uniform/stratified samples and store them. Then at run-time, they execute the queries against these stored samples thus reducing query latency. In order to reduce error for different set of queries --- using different columns for filters and groupings --- they need to create not one, but multiple sets of stratified samples of the same dataset. This results in a significant storage overhead.
Recently, following the promising results~\cite{aischeduling, mirhoseini2017device, kraska2018case} of using Machine Learning (ML) to solve several systems problems, few ML-based AQP techniques have been proposed. These use either special data-structures (DeepDB~\cite{deepdb}) or simple generative models (VAE-AQP~\cite{vae}) to answer queries with lower latency.

\begin{figure}
  \begin{minipage}{1.0\columnwidth}
    \centering
    \includegraphics[width=0.7\linewidth]{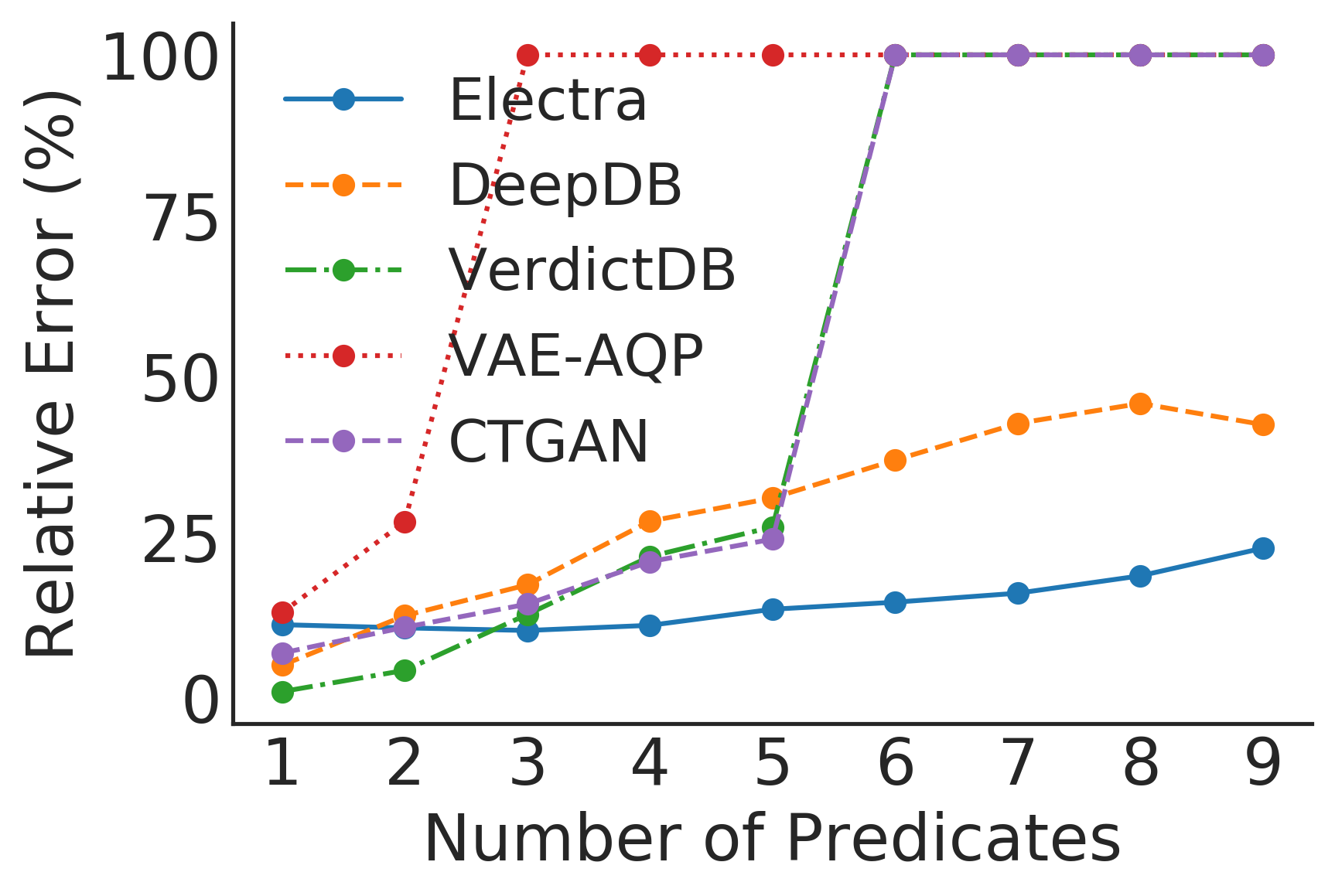}
    \caption{Median relative error vs. no. of predicates. Baselines have high error for important high-predicate queries.}
    \label{fig:rel_error_num_predicates}
  \end{minipage}
\end{figure}

However, a key drawback of these techniques is that as the number of predicates (i.e. \texttt{WHERE} conditions) in the query increases, the approximation error significantly increases (Fig.~\ref{fig:rel_error_num_predicates} for Flights dataset). \name is our technique shown in \textit{blue}.
The importance of the queries containing such large number of predicates is disproportionally high.
This is because, as expert analysts drill down for insights, they progressively use more and more predicates to filter the data to arrive at the desired subset and breakdowns where some interesting patterns might be present.
Therefore, lowering the approximation error for such high-predicate queries can prevent wrong conclusions, after laborious dissection of data in search of insights. As shown in Fig.~\ref{fig:rel_error_num_predicates}, all baselines fail to achieve this goal and therefore can be detrimental for complex insight discovery tasks.

In this paper, we present \name, a predicate-aware AQP system that can answer analytics-style queries having a large number of predicates with much lower approximation errors compared to the state-of-the-art techniques.
The key novelty of \name comes from the use of a \textit{predicate-aware generative model} for AQP and associated training methodology to capture accurate conditional distribution of the data even for highly selective queries. \name then uses the conditional generative model~\cite{cvae} to generate a few representative samples at runtime for providing an accurate enough answer.

The attributes (columns) in a tabular dataset can have categories (groups) that are rare. 
Moreover, when a large number of predicates are used to filter the data, the number of rows in the original data satisfying such conditions can be very small.
If we cannot faithfully generate such rare categories with proper statistical characteristics, AQP results would either have large errors or even miss entire groups for queries using a \texttt{GROUP BY}.
To overcome this problem, we propose a \textit{stratified masking strategy} while training the CVAE and an associated modification of the CVAE loss-function to help it learn predicate-aware targeted generation of samples. Specifically, using this proposed masking strategy, we help the model to learn rare groups or categories in the data.
Our empirical evaluations on real-world datasets show that, our stratified masking strategy helps \name to learn the \textit{conditional distribution} among different attributes, in a query-agnostic manner and can provide an average 12\% improvement in accuracy (Fig.~\ref{fig:masking_results}) compared to \textit{random} masking strategy proposed by prior work~\cite{vaeac} (albeit, not for AQP systems but for image generation).
Our major contributions in this paper are:
\begin{itemize}
\item We propose a novel and low-error approximate query processing technique using predicate-aware generative models. Even for queries with a large number of predicates, our technique can provide effective answers by generating a few representative samples, at the client-side.
\item We propose a novel \textit{stratified-masking} technique that enables training the generative model to learn better conditional distribution and support predicate-aware AQP resulting in a significant performance improvement on low selectivity queries.
\item We present the end-to-end design and implementation details of our AQP system called \name. We present detailed evaluation and ablation study with real-world datasets and compare with state-of-the-art baselines. Our technique reduces AQP error by 36.6\% on average compared to previously proposed generative-model-based AQP technique on production workloads.
\end{itemize}
\section{Overview of \name}

\subsection{Target Query Structure}
\label{sec:goal}
The commonly used queries for interactive data exploration and visual analytics have the following structure:
\newline
{\footnotesize
\noindent\texttt{\textbf{SELECT} $\Psi$, \textbf{AGG}($N$) as $\alpha$ \textbf{FROM} $T$ \textbf{WHERE} $\theta$ [\textbf{GROUP BY} $C$] [\textbf{ORDER BY} $\beta$] 
}
}
\normalsize

Aggregate AGG (\texttt{AVG}, \texttt{SUM} etc.) is applied to a numerical attribute $N$, on rows satisfying predicate $\theta$. \texttt{GROUP BY} is optionally applied on categorical attributes $C$.
One such query example is shown in Fig.~\ref{fig:runtime_electra}.
Here an analyst might be exploring what are the average checkout values made by users from a certain demographic (Age ``Senior" using ``iOS" and ``Chrome" Browser) from an e-commerce site across different months. Insights like these helps organizations to understand trends, optimize website performance or to place effective advertisements.
\name can handle a much diverse set of queries including queries with multiple predicates by - breaking down the query into multiple queries; using inclusion-exclusion principle or converting the query's predicates to Disjunctive Normal Form.
For more details, please refer to the supplement section \textit{Query Handling}.

\noindent\textbf{Predicates:}
Predicates are condition expressions (e.g., \texttt{WHERE}) evaluating to a boolean value used to filter rows on which an aggregate function like \texttt{AVG}, \texttt{SUM} and \texttt{COUNT} is to be evaluated. A compound predicate consists of multiple (\texttt{AND}s) or (\texttt{OR}s) of predicates.

\noindent\textbf{Query Selectivity:}
The \textit{selectivity} of a query with predicate $\pi$ is defined as
$sel\left(\pi\right) \coloneqq \lvert{x \in T  \colon \pi\left(x\right) = 1}\rvert / \lvert{T}\rvert$. It indicates what fraction of rows passed the filtering criteria used in the predicates.

\subsection{Design of \name}
\label{sec:design}
\label{sec:sys_design}
\begin{figure}[ht]
\begin{subfigure}[b]{1.0\columnwidth}
\includegraphics[width=\linewidth]{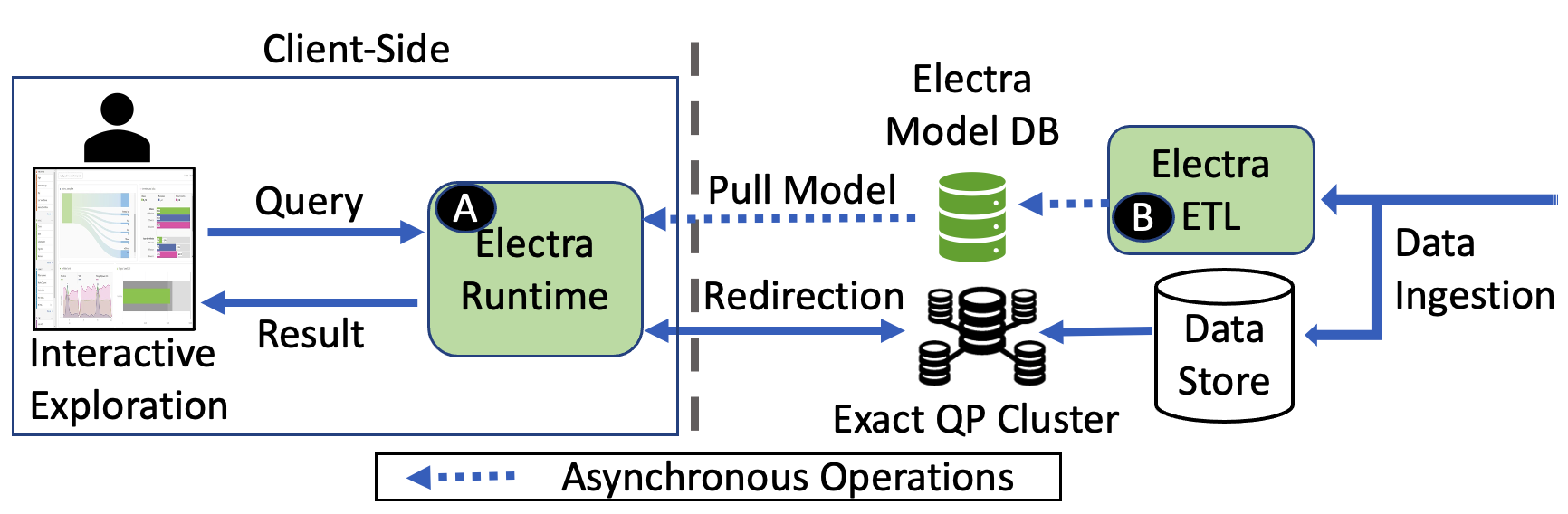}
\caption{Overall architecture of \name}
\label{fig:overview_deployment}
\end{subfigure}
\begin{subfigure}[b]{1.0\columnwidth}
\includegraphics[width=0.9\linewidth]{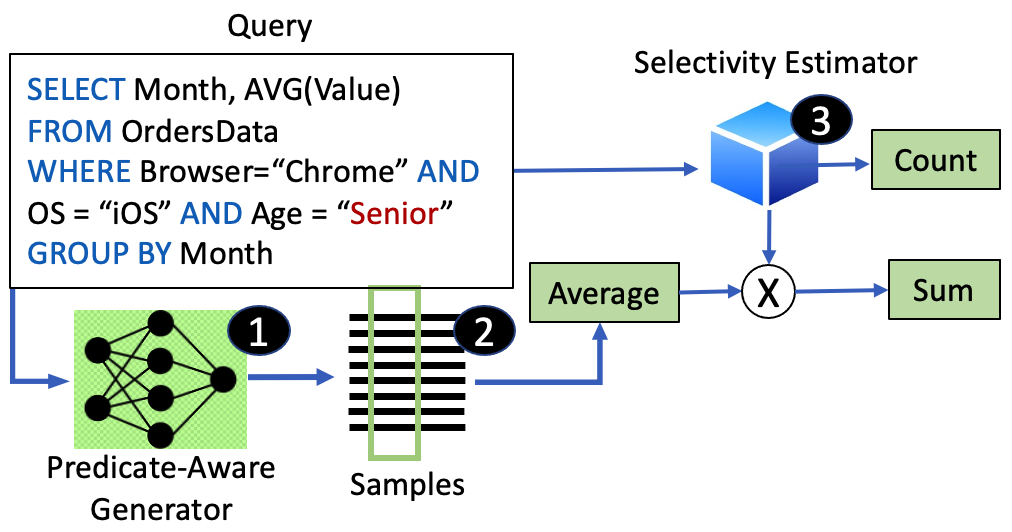}
\caption{\name Runtime computations}
\label{fig:runtime_electra}
\end{subfigure}
\caption{Design of \name}
\label{fig:deployment}
\end{figure}

Fig.~\ref{fig:overview_deployment} shows the major components of \name (marked in green).
At a high-level, it has primarily two components: \circled{A} an on-line AQP runtime at the client-side and, \circled{B} an offline server-side ETL (Extraction, Transformation, and Loading) component.
The runtime component \circled{A} is responsible for answering the live queries from the user-interactions using the pre-trained ML-models. It pulls these models asynchronously from a repository (Model-DB) and caches them locally.
\name is designed to handle the most frequent-type of queries for exploratory data analytics (Refer \S~\ref{sec:goal}). Unsupported query-formats are \textit{redirected} to a backend query-processor.

Fig.~\ref{fig:runtime_electra} shows more details of the runtime component. 
\name first parses the query to extract individual predicates, \texttt{GROUP BY} attributes, and the aggregate function.
If the aggregate is either a \texttt{SUM} or an \texttt{AVG}, \name feeds the predicates and the \texttt{GROUP BY} information to its predicate-aware ML-model (\circled{1}) and generates a small number of representative samples.
For queries with \texttt{AVG} as an aggregate, \name can directly calculate the result by applying the query logic on these generated samples (\circled{2}).
For \texttt{SUM} queries, \name needs to calibrate the result using an appropriate scale-up or denormalizing factor. 
\name calculates this factor by feeding the filtering predicates to a \textit{selectivity estimator} (\circled{3}).
For \texttt{COUNT} queries, \name directly uses the selectivity estimator to calculate the query result.

The server-side ETL component (\circled{B} in Fig.~\ref{fig:overview_deployment}) operates asynchronously.
For a new dataset, it trains two types of models -- a conditional generative model and a selectivity estimator model -- and stores them into the Model-DB.

\section{Conditional Sample Generation}
\label{sec:model_details}
In this section, we describe how \name attempts to make conditional sample generation process accurate, even for queries with large number of predicates, so that the resulting approximation error on these samples becomes very small.

\subsection{Problem Description}
Let T be a relation with $K = |A|$ attributes where $A = A_N \cup A_C$. $A_N$ denotes the numerical attributes and $A_C$ denotes the categorical attributes. Our aim is to learn a model that can generate a predicate-aware representative sample of T at runtime. 
This problem can be divided into two sub-problems -- first, learning the conditional distributions of the form $P(A_n|A_c)$ where $A_n$ is a subset of $A_N$ and $A_c$ is a subset of $A_C$ and second, generating representative samples using the learned conditional distributions.
With respect to the example query in Fig.~\ref{fig:runtime_electra}, the attribute \texttt{Value} belongs to $A_N$. The attributes \texttt{Browser}, \texttt{OS}, \texttt{Age} and \texttt{Month} used in the predicates are categorical attributes and belongs to $A_C$.

\name targets to support queries with arbitrary number of predicate combinations over the attributes $A_C$ using only a single model. This requires the learning technique to be \textit{query-agnostic} (but yet, sample generation to be \textit{query-predicate specific}).

\subsection{\name's Model Design}
We are the first to propose the use of Conditional Variational Auto-encoders (CVAE)~\cite{cvae} along with a query-agnostic training methodology involving a novel masking strategy to reduce the AQP error.

Consider a $K$-dimensional vector $a$ with numerical and categorical values representing a row of data. Let $m \in {\{0,1\}}^{K}$ be a binary mask denoting the set of unobserved ($m_{i} = 1$) and observed attributes ($m_{i} = 0$). Applying this mask on the vector $a$ gives us $a_m = \{a_{i:m_i = 1}\}$ as the set of unobserved attributes and $a_{1-m} = \{a_{i:m_i=0}\}$ as the set of observed attributes.
Our goal is to model the conditional distribution of the unobserved $a_m$ given the mask $m$ and the observed attributes $a_{1-m}$ i.e. approximate the true distribution $p(a_{m}|a_{1-m},m)$.

Along with providing full support over the domain of $m$ (arbitrary conditions), we need to ensure that the samples that are generated represent rows from the original dataset. For example: given a set of observed attributes $a_{1-m}$ whose combination does not exist in the data, the model should not generate samples of data satisfying $a_{1-m}$ (since a \textit{false positive}). To avoid this, we learn to generate complete $a$ (both $a_{m}$ and $a_{1-m}$) by learning the distribution $p(a|a_{1-m},m)$ instead of just learning $p(a_{m}|a_{1-m},m)$ and hence avoiding generating \textit{false positives}.

Note that the generative models based on VAE(s)~\cite{aevb} allows us to generate random samples by generating latent vector $z$ and then sampling from posterior $p_{\theta}(x|z)$. But, we do not have any control on the data generation process. Conditional Variational Auto-encoder models (CVAE)~\cite{cvae} counter this by conditioning the latent variable and data generation on a control variable. In terms of the table attributes $a$, the control variable $a_{1-m}$ and the mask $m$, the training objective for our model thus becomes:

\useshortskip
{\small
\begin{multline}
\mathcal{L_{\text{\name}}} = -KL( q_{\phi}(z|a,m)||p_{\psi}(z|a_{1-m},m)) + \\
\mathbb{E}_{z \sim q_{\phi}(z|a,m)}[log(p_{\theta}(a|z))]
\end{multline}
}
\normalsize

The loss function is a combination of two terms: (1) KL-divergence of the posterior and the conditional distribution of the latent space i.e. how well is your conditional latent space generation and (2) Reconstruction error for the attributes i.e. how well is your actual sample generation.

The generative process of our model thus becomes: (1) generate latent vector conditioned on the mask and the observed attributes: $p_{\psi}(z|a_{1-m},m)$ and (2) generate sample vector $a$ from the latent variable $z$: $p_{\theta}(a|z)$, thus inducing the following distribution:

\useshortskip
{\small
\begin{equation}
    p_{\psi,\theta}(a|a_{1-m},m) = \mathbb{E}_{z \sim p_{\psi}(z|a_{1-m},m)} p_{\theta}(a|z)
\end{equation}
}
\normalsize
Since $a_m$ and $a_{1-m}$ can be variable length vectors depending on $m$, inspired from \cite{vaeac}, we consider $a_m = a \circ m$ i.e. the element-wise product of $a$ and $m$. Similarly, $a_{1-m} = a \circ (1-m)$.

In terms of a query, the observed attributes are the predicates present in the query. 
E.g. for the query in Fig.~\ref{fig:runtime_electra}, for attributes $A$ = \texttt{\{Month, Value, Browser, OS, Age\}}, $m$ is $\{1,1,0,0,0\}$ and the observed attributes vector $a_{1-m}$ is \texttt{\{-, -, "Chrome", "iOS", "Senior"\}}.

\subsubsection{Model Architecture:}
\begin{figure*}[t]
\begin{subfigure}[t]{0.24\textwidth}
  \includegraphics[width=1.0\linewidth]{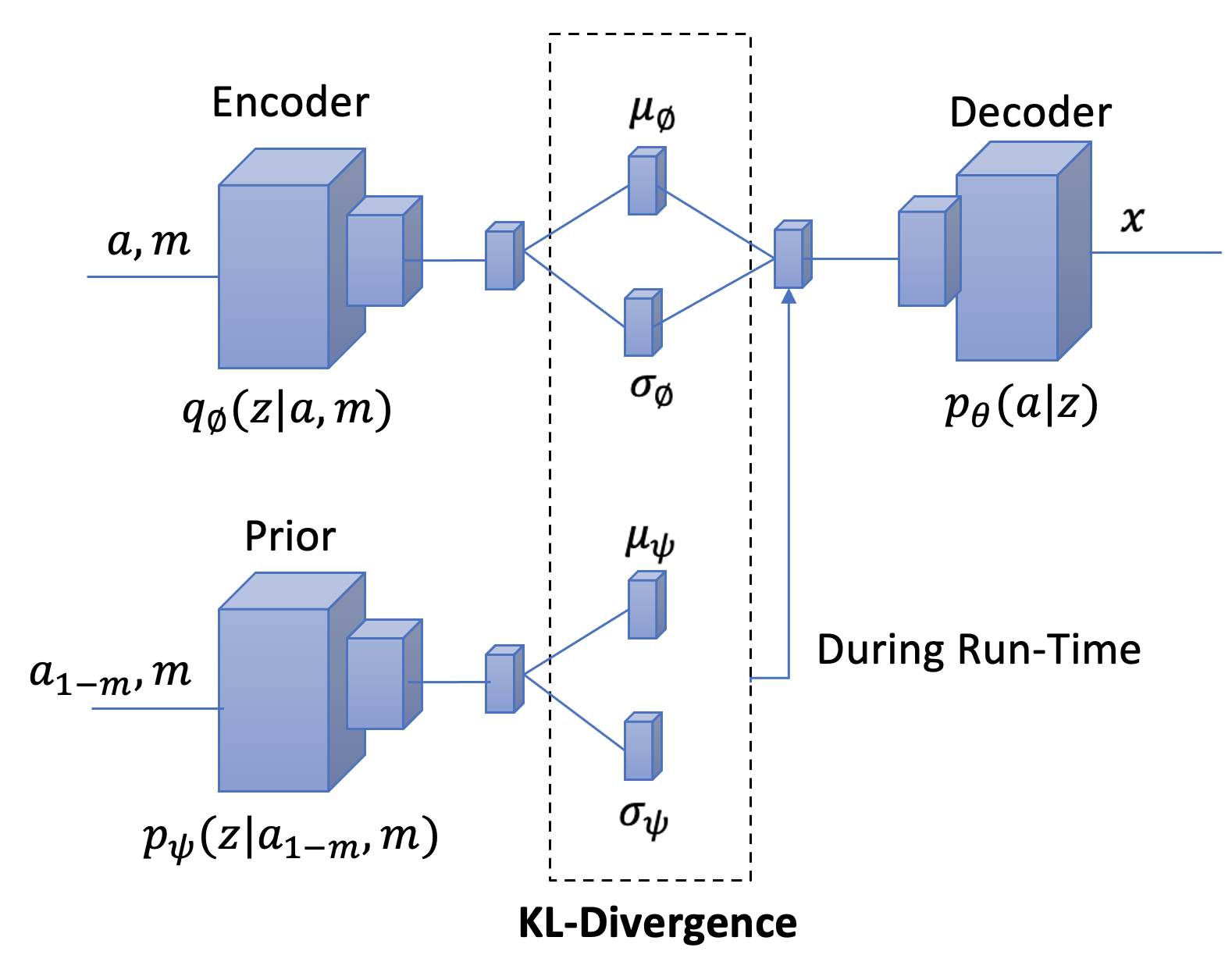}
  \caption{Model architecture}
  \label{fig:model_architecture_diagram}
\end{subfigure}%
\hfill
\begin{subfigure}[t]{0.45\textwidth}
  \includegraphics[width=1.0\linewidth]{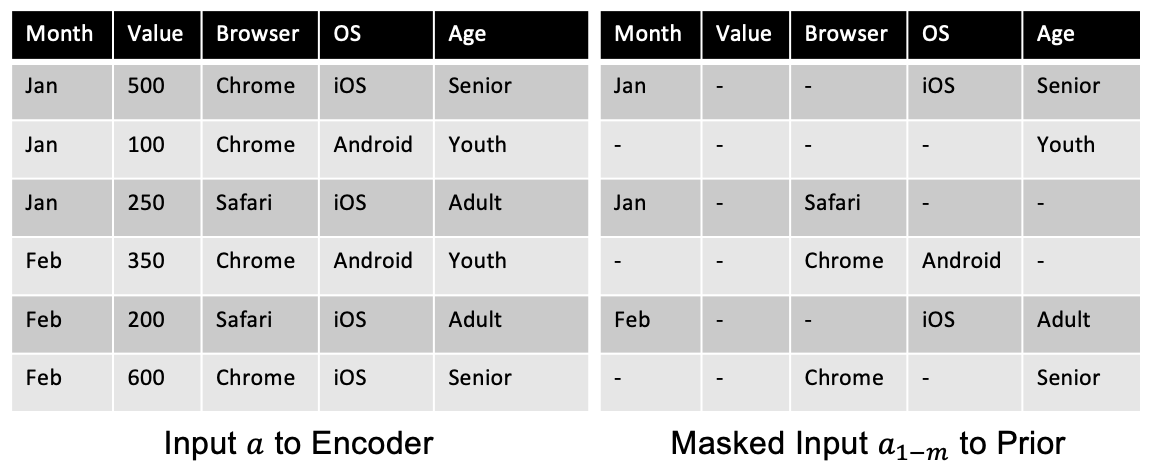}
  \caption{Training inputs}
  \label{fig:model_architecture_train_input}
\end{subfigure}%
\hfill
\begin{subfigure}[t]{0.22\textwidth}
  \includegraphics[width=1.0\linewidth]{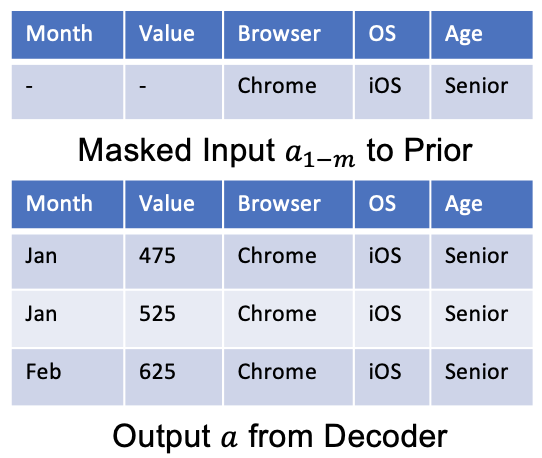}
  \caption{Runtime inputs}
  \label{fig:model_architecture_test_input}
\end{subfigure}%
\caption{Model architecture and inputs}
\label{fig:model_architecture}
\end{figure*}

~\name's CVAE model consists of three networks - (1) Encoder: to approximate the true posterior distribution, (2) Prior: to approximate the conditional latent space distribution, and (3) Decoder: to generate synthetic samples based on the latent vector $z$. We use a Gaussian distribution over $z$ with parameters $(\mu_{\phi},\sigma_{\phi})$ and $(\mu_{\psi},\sigma_{\psi})$ for the encoder and prior network respectively.
Fig.~\ref{fig:model_architecture_diagram} shows the model architecture. All the three networks are used during training, whereas during runtime, only the prior and decoder networks are used.
Fig.~\ref{fig:model_architecture_train_input} shows the inputs to these networks during training, corresponding to our example dataset mentioned in Fig.~\ref{fig:runtime_electra}.
The input to the Prior network is masked using a novel masking strategy. Fig.~\ref{fig:model_architecture_test_input} illustrates the runtime inputs to \name's generative model for our example query. It also illustrates the generated samples as obtained from the Decoder network, corresponding to the input query.
After generation of such targeted samples that preserve the conditional distribution of the original data in the context of the query received at runtime, \name simply executes the original query on this very small sample to calculate an answer.

\subsection{Stratified Masking}
\label{sec:masking}
The efficiency of learned conditional distributions depends on the set of masks $M = \{m\}$ used during the training. During each epoch, a mask $m_{b \times k}$ is generated for a batch of data. This mask is applied to the input data to generate the set of observed attributes, for which the samples are generated and the loss function evaluated.
In case of $m = {0}^{K}$ i.e. no masking and all the attributes are observed, the prior network would learn to generate latent vectors corresponding to all observed attributes. But at runtime, only a partial set of attributes are observed (as predicates) and hence it would lead to a poor performance.
In case of $m = {1}^{K}$ i.e. a masking rate of 100\% and all the features are unobserved, the behavior would be similar to that of a VAE. Since the vector $a_{1-m} = \{\phi\}$, the conditioning of the latent vector would have no effect on the learning. Thus, again this would perform poorly on queries with predicates.

In order to counter these and make the training query-agnostic, a random masking strategy~\cite{vaeac} can be used where a certain fraction of the rows are masked corresponding to each of the input features.
However, such random masking does not provide any mechanism to help the model learn better the conditional distribution for rare groups to improve performance for queries with large number of predicates.
There are more subtleties. By masking more number of rows for an attribute, the training helps the model to learn better the generation of representative values for that attribute. Whereas, by keeping more unmasked rows, the model can learn better conditional distribution conditioned on those attributes.

{\small
\begin{algorithm}[ht]
\SetKwInput{KwInput}{Input}
\SetKwInput{KwOutput}{Output}
\SetAlgoLined
\KwInput{Batch $\{B\}_{b \times k}$, Masking Factor $r$, Strata Size $S$}
\KwOutput{Binary mask $m$ for the Input Batch $B$}
$m \gets {0}_{b \times k}$ \\
$num\_cols \gets numerical\_cols(k)$ \\
$cat\_cols \gets categorical\_cols(k)$ \\
\For{$i$ in $num\_cols$}{
    $\{m_{x,y}\}_{1,i}^{b,i} \gets 1$ \\
}
\For{$i \in cat\_cols$}{
    $values \gets \{B_{x,y}\}_{1,i}^{b,i}$ \\
    $weights \gets \{S_{i}[values_{x}]\}_{1}^{b}$ \\
    $batch\_indices \gets random.sample(\{x\}_{1}^{b}, r*b, weights)$ \\
    \For{$j \in batch\_indices$}{
        $m_{j,i} \gets 1$ \\
    }
}
\Return{$m;$}
\caption{Stratified Masking Strategy}
\label{algo:stratified_masking}
\end{algorithm}
}

\noindent\textbf{Proposed masking strategy: }
With the above observation, \name uses a novel masking strategy tailored for AQP.
It consists of the following:
(a) We completely mask the numerical attributes.
(b) For the categorical attributes on which conditions can be used in the predicates, we use a \textit{stratified masking strategy}.
In the stratified masking we alter the random masking strategy based on the size of the strata (i.e. groups or categories) to disproportionately reduce the probability of masking for the rare groups. Thus, the ability of the model to learn better conditional distribution improves.
We apply this masking strategy in a batch-wise manner during training. In the Evaluation section, we show (Fig.~\ref{fig:masking_results}) that our masking strategy drastically reduces the approximation error compared to no-masking or random masking strategies. %

Algorithm \ref{algo:stratified_masking} describes the masking strategy where we input a batch of data $B$ of dimension $b \times k$: $b$ denotes the batch size and $k$ denotes the number of attributes. The numerical columns are completely masked. The strata size for each categorical column $i$ ($S_i$), contains for each value that the column $i$ can take, the fraction of rows in the data where the column takes this value. Using the $weights$ obtained from $S_i[value]$ as the sampling probability and $r$ as the masking factor, we sample $r * b$ rows and mask their column $i$.
Masking factor $r$ controls the overall masking rate of the batch. A higher masking factor simulates queries with less observed attributes i.e. less number of predicates and a lower masking factor simulates queries with more observed attributes i.e. more number of predicates.
\section{Implementation Details}
\label{sec:implementation}

\noindent\textbf{Data Transformation.}
\name applies a set of data pre-processing steps including label encoding for categorical attributes and mode-specific normalization for numerical attributes. For normalization, weights for the Gaussian mixture are calculated through a variational inference algorithm using sklearn's \texttt{BayesianGaussianMixture} method. The parameters specified are the number of modes (identified earlier) and a maximum iteration of 1000. To avoid mixture components with close to zero weights, we limit the number of modes to less than or equal to 3.

\noindent\textbf{Conditional Density Estimator.}
The conditional density estimator model is implemented in PyTorch. We build upon the code~\footnote{VAEAC code available at https://github.com/tigvarts/vaeac} provided by Ivanov et al.
The density estimator is trained to minimize the combination of KL-divergence loss and reconstruction loss. To select a suitable set of hyperparameters, we performed a grid search. We varied the depth ($d$) of the prior and proposal networks in the range [2,4,6,8] and the latent dimension ($L$) in the range [32,64,128,256]. For Flights data we use $d = 8$, $L = 64$, for Housing $d = 8$, $L = 64$, and for Beijing PM2.5 we use $d = 6$, $L = 32$.  Note that, the depth of the networks and the latent dimension contribute significantly to the model size. Hence, depending on the size constraints (if any), one can choose a simpler model. We used a masking factor ($r$) of 0.5. The model was trained with an Adam Optimizer with a learning rate of 0.0001 (larger learning rates gave unstable variational lower bound(s)).

\noindent\textbf{Selectivity Estimator.}
We use \naru's publicly available implementation~\footnote{\naru code available at https://github.com/naru-project/naru/}. The model is trained with the ResMADE architecture with a batch size of 512, an initial warm-up of 10000 rounds, with 5 layers each of hidden dimension 256. In order to account for the impact of \textit{column ordering}, we tried various different random orderings and chose the one with the best performance.

\textit{We provide more details about the pre-processing steps and training methodology in the supplement section:  Implementation Details.}
\section{Evaluation}
\label{sec:eval}

\begin{figure*}[t]
\centering
\begin{minipage}[t]{0.62\textwidth}
\vspace{0pt}
\begin{subfigure}[t]{1.0\textwidth}
\includegraphics[width=\textwidth]{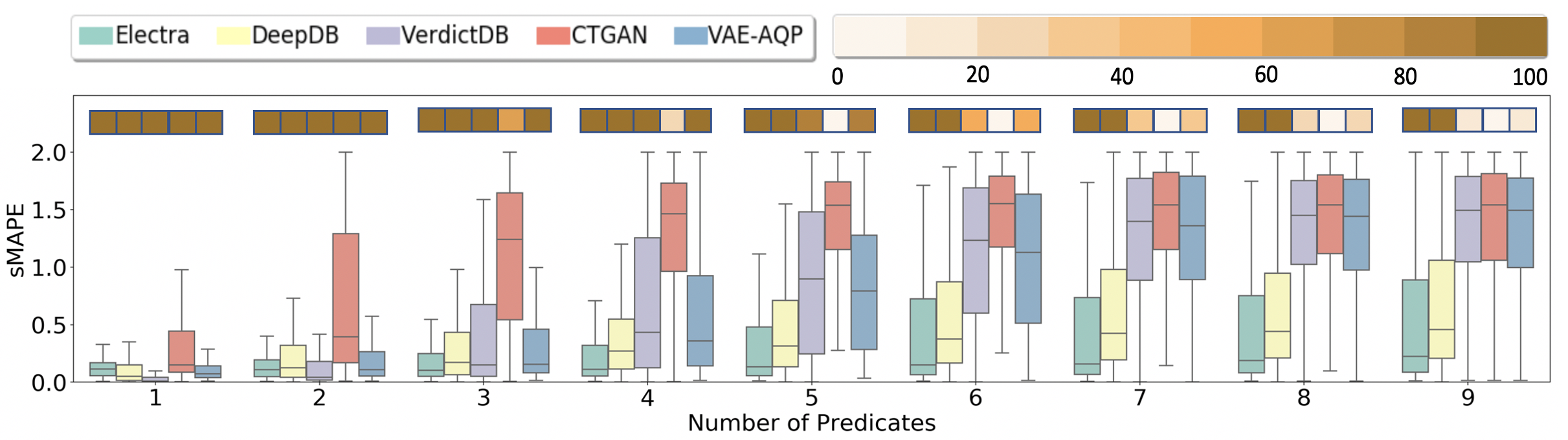}
\caption{Flights}
\end{subfigure}
\hfill
\begin{subfigure}[t]{1.0\textwidth}
\includegraphics[width=1.0\textwidth]{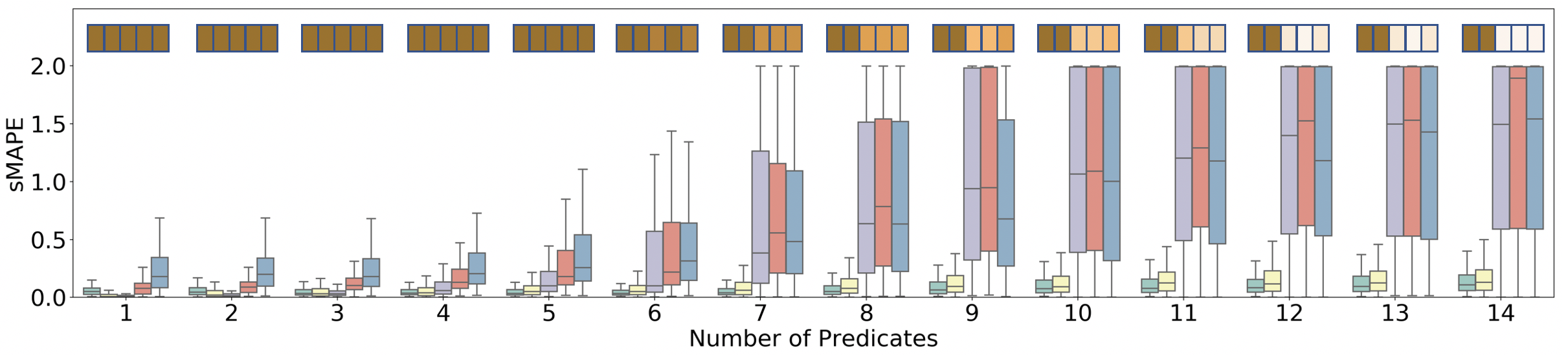}
\caption{Housing}
\end{subfigure}
\hfill
\begin{subfigure}[t]{1.0\textwidth}
\includegraphics[width=1.0\textwidth]{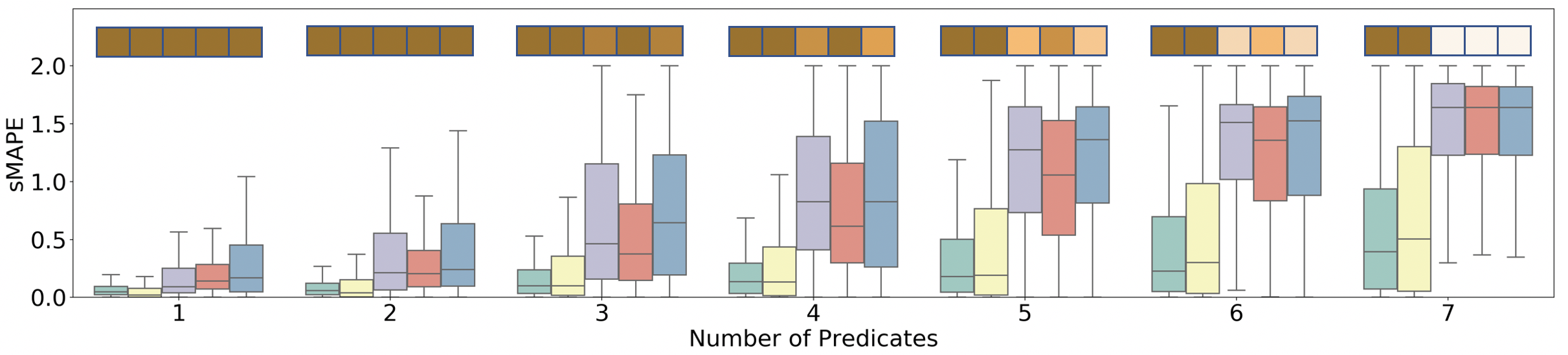}
\caption{Beijing PM2.5}
\end{subfigure}
\caption{AQP error vs. \# predicates. Heat-map shows \% of queries answered.}
\label{fig:smape-box}
\end{minipage}%
\hfill
\begin{minipage}[t]{0.35\textwidth}
\vspace{0pt}
\begin{minipage}[t]{1.0\textwidth}
\begin{subfigure}[t]{1.0\textwidth}
\includegraphics[width=\textwidth]{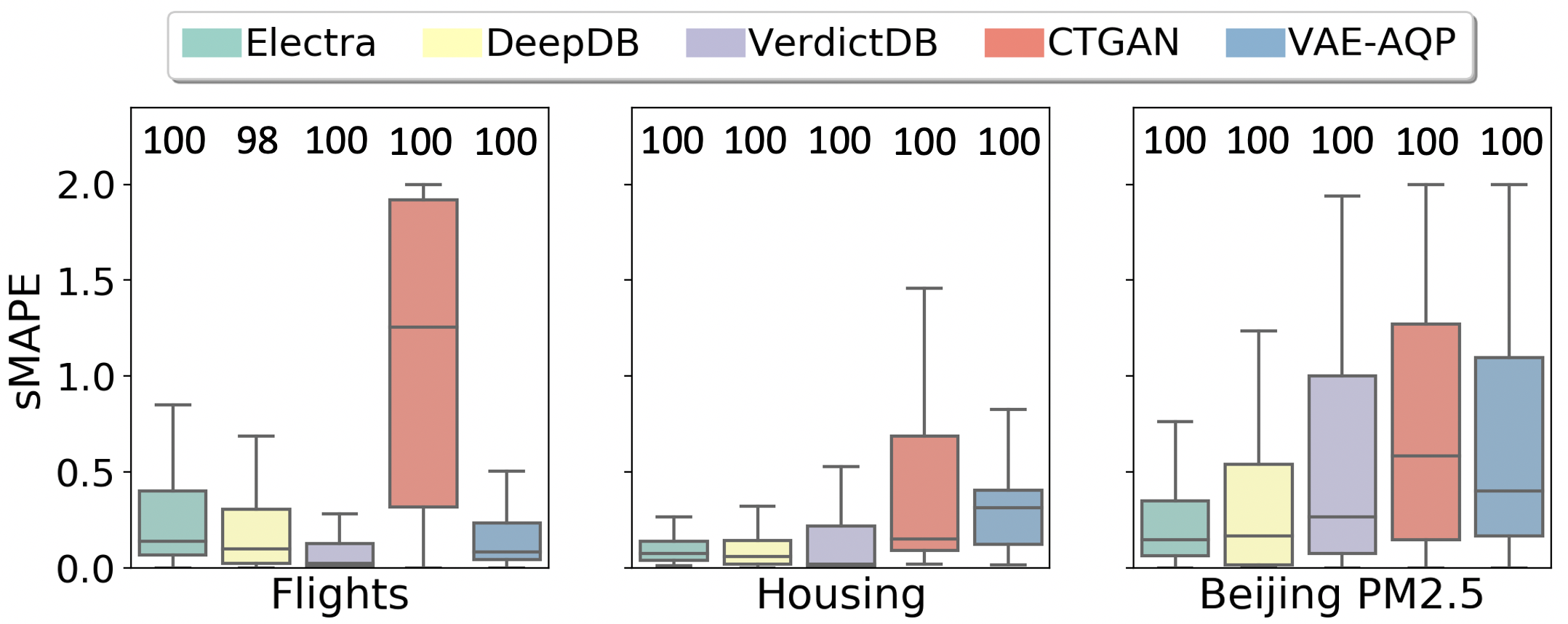}
\caption{AVG queries}
\label{fig:aep-workload-avg}
\end{subfigure}
\hfill
\begin{subfigure}[t]{1.0\textwidth}
\includegraphics[width=\textwidth]{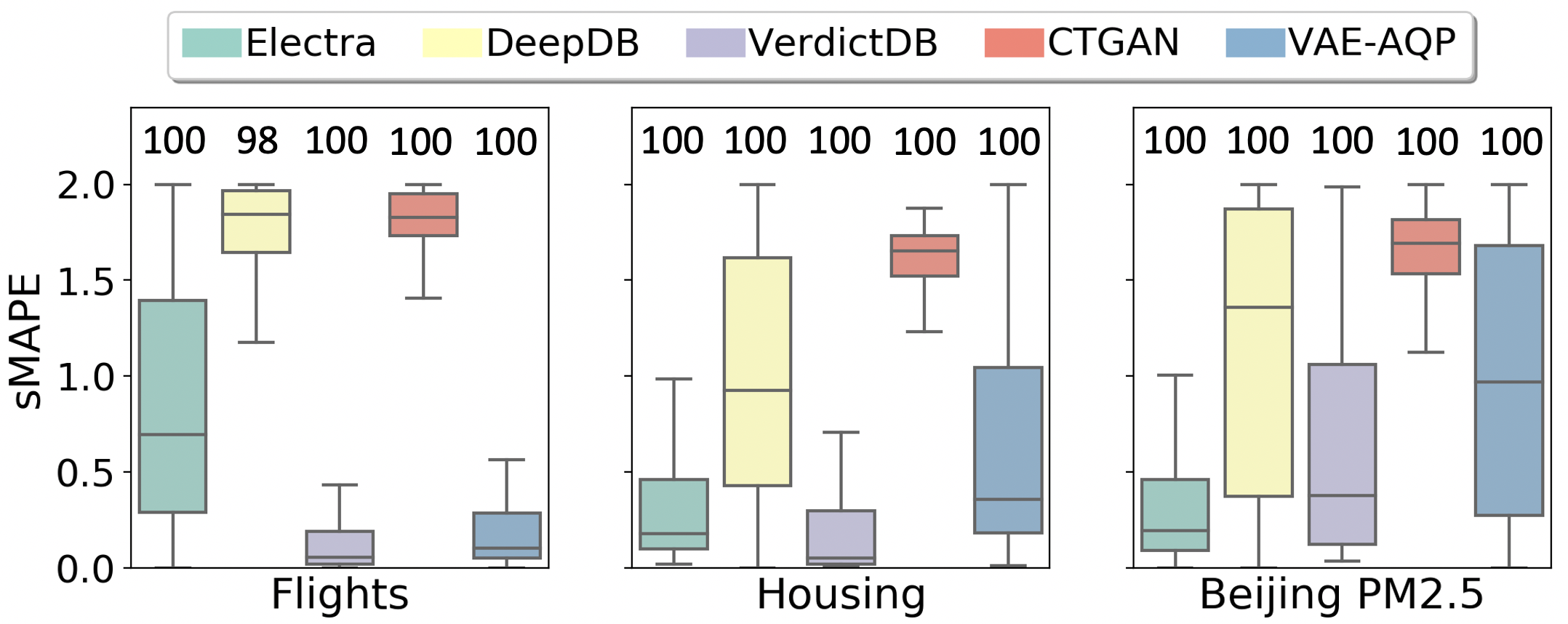}
\caption{SUM queries}
\label{fig:aep-workload-sum}
\end{subfigure}
\caption{AQP error vs. \# predicates. number on top shows \% of queries answered.}
\label{fig:aep-workload}
\end{minipage}
\hfill
\begin{minipage}[t]{1.0\textwidth}
\scalebox{0.82}{
\begin{tabular}[t]{>{\itshape} l|c|c|c|c}
    \toprule
    \multirow{1}{*}{Dataset} & \multicolumn{4}{c}{Sample size} \\
    \cline{2-5}
    & 500 & 1000 & 1500 & 2000 \\
    \midrule
    {Flights} & 15.15 & 14.93 & 14.70 & \textbf{14.60} \\
    {Housing} & 5.96 & \textbf{5.91} & 5.92 & 5.92 \\
    {Beijing PM2.5} & 15.98 & 15.79 & 15.77 & \textbf{15.64} \\
    \midrule
    \end{tabular}
    }
\captionof{table}{Median R.E. vs. \# samples generated}
\label{tab:num-samples-variation}
\end{minipage}%
\end{minipage}
\end{figure*}

\subsection{Experimental Environment}
All the experiments were performed on a 32 core Intel(R) Xeon(R) CPU E5-2686 with 4 Tesla V100-SXM2 GPU(s).

\noindent\textbf{Datasets:}
We use three real-world datasets: Flights~\cite{Flights}, Housing~\cite{housing} and Beijing PM2.5~\cite{BeijingPM25}.
The Flights dataset consists of arrival and departure data
(e.g. carrier, origin, destination, delay) for domestic flights in the USA.
The Housing dataset consists of housing price records (e.g. price, area, rooms) obtained from a real-estate company. 
The Beijing PM2.5 dataset consists of hourly records of PM2.5 concentration.
All these datasets have a combination of both categorical and numerical attributes.
(Ref. Supplement Table 1).

\noindent\textbf{Synthetic Query Workload:}
We create a synthetic workload to evaluate the impact of the number of predicates in the query. For each dataset, we generated queries with \textit{\# of predicates} ($k$) ranging from 1 to $|A_C|$. For a given $k$, we first randomly sample 100 combinations (with repetitions) from the set of all possible $k$-attribute selections. Then, for each of these selected $k$-attributes, we create the \texttt{WHERE} condition by assigning values based on a randomly chosen tuple from the dataset. Then, we use \texttt{AVG} as the aggregate function for each of the numerical attributes. The synthetic workload thus contains a total of $100*|A_C|*|A_N|$ queries.

\noindent\textbf{Production Query Workload:}
We also evaluate \name using real production queries from five Fortune 500 companies over a period of two months on a large data exploration platform. First, we select queries whose query structure is supported by \name. We filter out the repeated queries. While we were able to extract the queries, we did not have access to the actual customer data corresponding to those queries. Hence, we use the following methodology to create an equivalent query set for our evaluation.
We replace each predicate in the \texttt{WHERE} clause by a random categorical attribute and a random value corresponding to that attribute. All attributes that were used in \texttt{GROUP BY} were replaced by random categorical attributes from our evaluation datasets. Then, we generate two workloads, one with \texttt{AVG} and one with \texttt{SUM} as the aggregate function on a random numerical attribute. Thus, we preserve real characteristics of the use of predicates and the grouping logic in our modified workload. 
\textit{For details on the query selectivity, please refer to the supplement.}

\subsubsection{Baselines:} We evaluate against four recent baselines.

\noindent\textbf{VAE-AQP}~\cite{vae} is a generative model-based technique that does not use any predicate information. The code was obtained from the authors. In our evaluations, we generated 100K samples from VAE-AQP for all 3 datasets. Note that this is well over the $1\%$ sampling rate specified by the VAE-AQP authors for these datasets.

\noindent\textbf{CTGAN}~\cite{ctgan} is a conditional generative model. However, the CTGAN technique as described in the paper ~\cite{ctgan} and the associated code~\cite{ctgan-code}, only support at most one condition for generation. Hence, when using CTGAN as a baseline in our evaluation, we randomly select one of the input predicates as the condition for generating the samples. We keep the generated sample size per query same as \name.

\noindent\textbf{DeepDB}~\cite{deepdb} builds a relational sum-product-network (SPN) based on the input data. For any incoming query, it evaluates the query on the obtained SPN network. We keep the default parameters for DeepDB code obtained from~\cite{deepdb-code}.

\noindent\textbf{VerdictDB}~\cite{verdictdb} is a sampling-based approach that runs at the server-side. It extracts and stores samples (called SCRAMBLE) from the original data and uses it to answer the incoming queries. We keep the default parameters in the code obtained from~\cite{verdictdb-code}.

\subsubsection{AQP Error Measures:}
\label{sec:error-measure}
For a query $Q$, with an exact ground truth result $g$ and an approximate result $a$, we measure the approximation error by computing the Relative Error (R.E.).

Since R.E. is an unbounded measure and can become very large, we use a bounded measure called Symmetric Mean Absolute Percentage Error (\textbf{sMAPE}) to visualize the approximation errors. sMAPE lies between 0 and 2 and provides a better visualization compared to R.E..

\useshortskip
{\small
\begin{align*}
    \textit{sMAPE}(Q) = 2*\frac{|g - a|}{|g|+|a|}
\end{align*}
}\normalsize

\subsection{Performance w.r.t. Number of Predicates}
Fig.~\ref{fig:smape-box} shows the distribution of approximation error (as sMAPE) w.r.t. the number of predicates in the synthetic workload queries. A lower approximation error is better.
All the queries could not be successfully evaluated to produce an approximate result because the technique could not match for the used predicates either in the generated data (\name, CTGAN, VAE-AQP) or in the available samples (VerdictDB) or in the stored metadata (DeepDB).
The distribution of the AQP errors in the box-plots were calculated only over the queries that the corresponding technique could support.
The heat-map colors over the box-plots show what \textit{percentages of the queries} were supported by different techniques. A technique supporting large fraction of queries is better.
It is worthwhile to note, only \name and DeepDB could answer the queries with $90-100\%$ coverage across all the different predicate settings.

Note, the median approximation error is almost always the lowest for \name (shown in green), as the number of predicates in the queries are increased. The other two generative baselines, CTGAN and VAE-AQP perform poorly in this aspect.
DeepDB's performance is  great for median approximation comparison, but error at the tail is often quite large for DeepDB, compared to \name. 
For one predicate, almost all the techniques can maintain very low AQP error (making any of those practically useful) and \name often does not provide the lowest error in the relative terms. However, \name is the overall best solution when we look at the whole spectrum of predicates. 

\begin{figure*}
\centering
\begin{minipage}[t]{0.495\textwidth}
\begin{minipage}[t]{0.49\textwidth}
\centering
\includegraphics[width=1.0\textwidth]{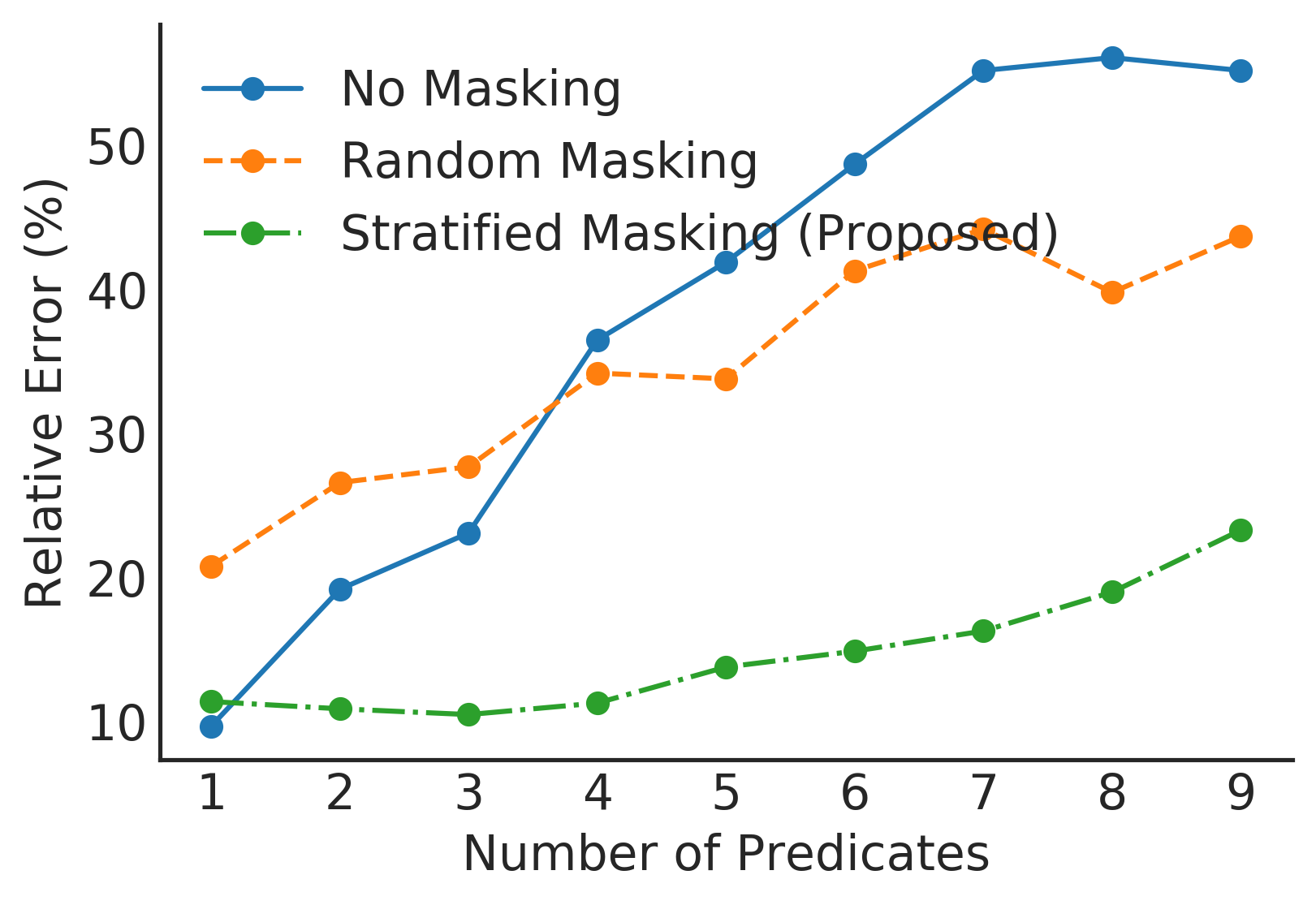}
\caption{Masking strategy}
\label{fig:masking_results}
\end{minipage}%
\hfill
\begin{minipage}[t]{0.49\columnwidth}
\includegraphics[width=1.0\textwidth]{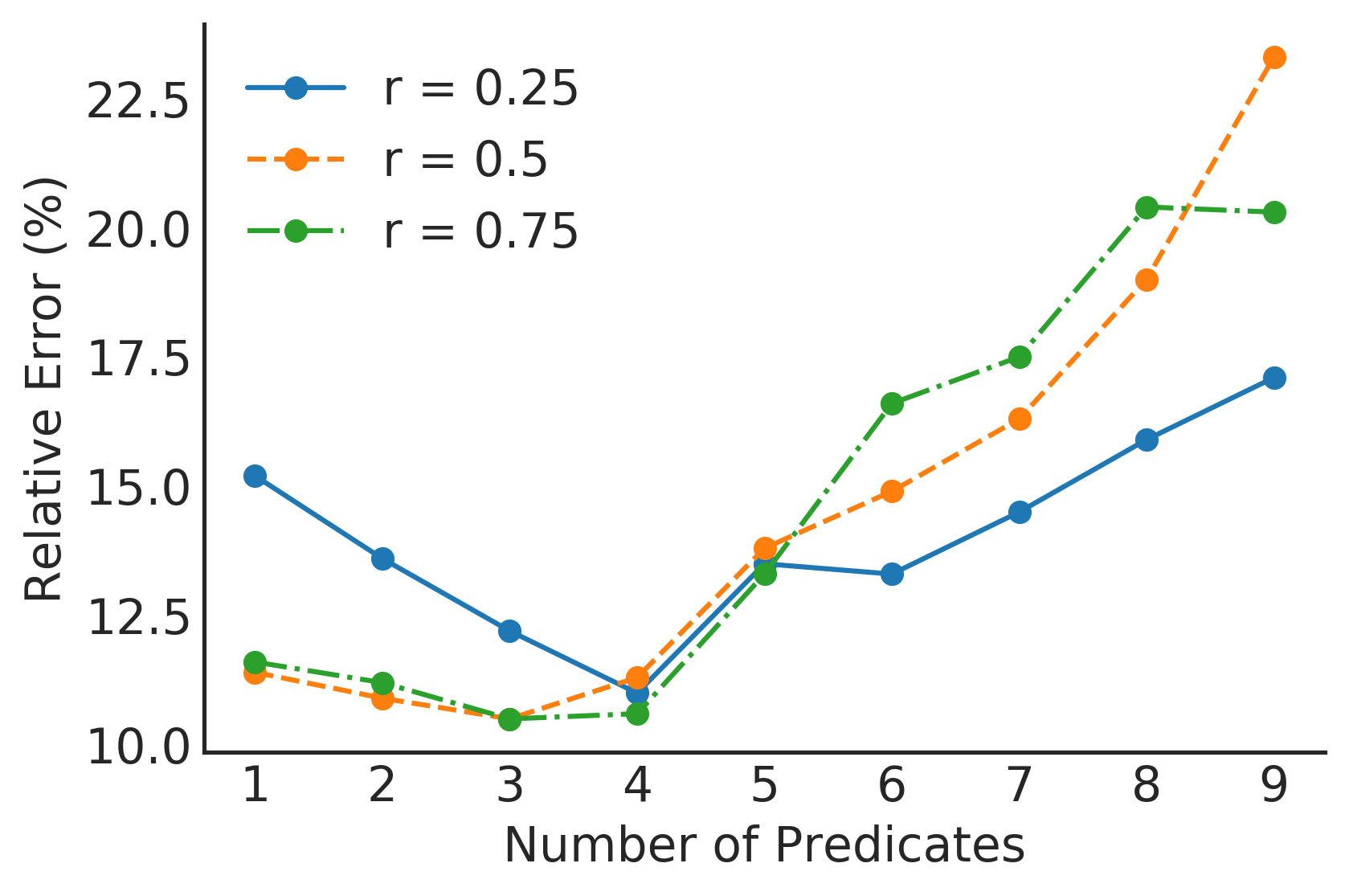}
\caption{Masking factor}
\label{fig:masking-factor}
\end{minipage}
\end{minipage}%
\hfill
\begin{minipage}[t]{0.48\textwidth}
   \includegraphics[width=\textwidth]{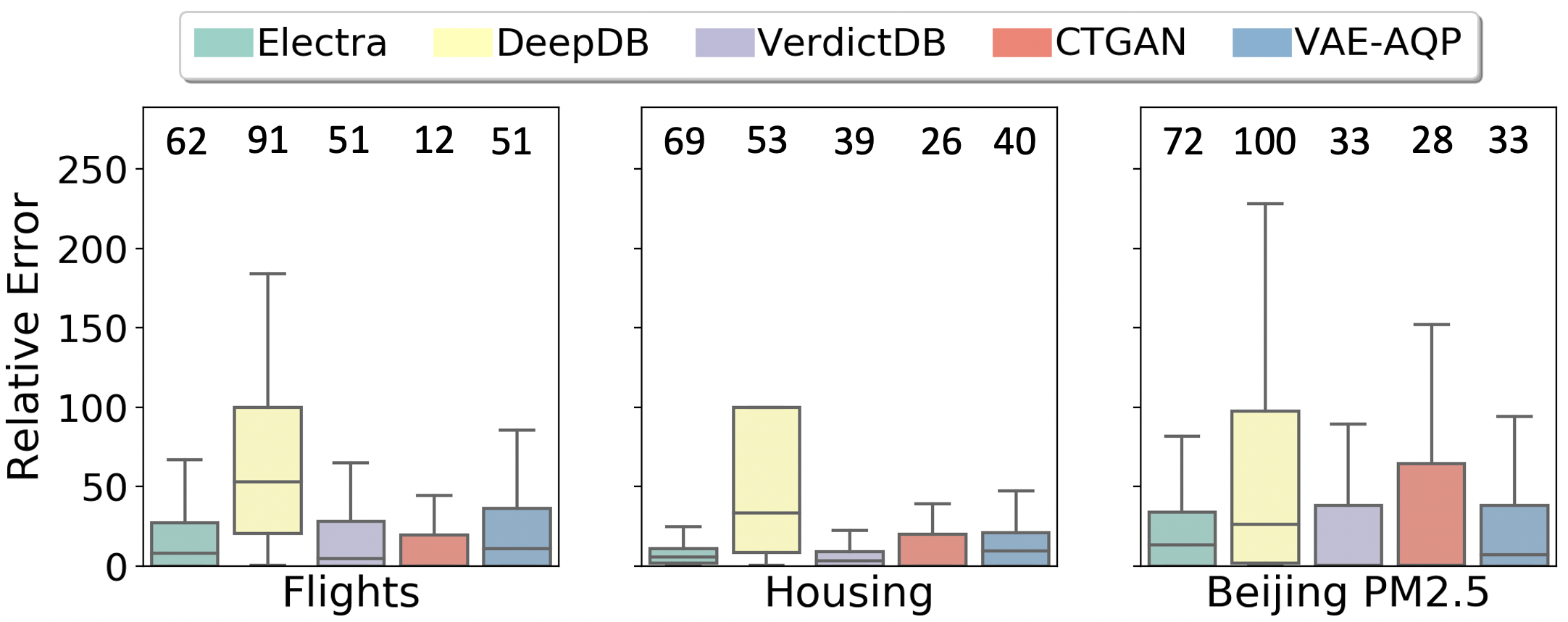}
    \caption{Error \& bin-Completeness for GROUP BY queries}
    \label{fig:group-by}
\end{minipage}
\end{figure*}

\subsection{Performance on Production Workload}
Fig.~\ref{fig:aep-workload} shows the comparisons of the AQP error (as sMAPE) distribution for our production query workload for the three real-world datasets shown separately for \texttt{AVG} and \texttt{SUM} as the aggregate functions. 
For AVG as the aggregate function (Fig.~\ref{fig:aep-workload-avg}), \name provides the lowest approximation error for both Housing and Beijing-PM-2.5 dataset. For Flights dataset, even though the median error is comparable to DeepDB, VerdictDB and VAE-AQP, the tail error is larger. 
Please note: since these plots were calculated over the production queries, 70\% of the queries had two or fewer predicates.
Thus, the benefit of \name in handling large number of predicates is less visible. The percentages of queries that different techniques were able to successfully answer, are shown over the box-plots. 

\subsection{Performance on \texttt{GROUP BY} Queries}

We evaluate \name's performance on \texttt{GROUP BY} queries using slight modification to our synthetic query workload by randomly choosing one of the attributes to be used for \texttt{GROUP BY}. We do not use any predicate on that attribute.
Fig.~\ref{fig:group-by} shows the AQP error over queries with all the different number of predicates. The error for each query was calculated as an average over the individual errors corresponding to each of the \textit{individual groups} that were both present in the ground truth results and the approximate results.
The error is computed as the average of the R.E. for each of the different groups corresponding to the \texttt{GROUP BY} column. 
Approximations might result in missing groups.
We use \textit{bin-completeness} measure as follows:
\useshortskip
{\small
\begin{align*}
    \textit{bin-completeness} = \frac{|G_{groups} \bigcap A_{groups}|}{|G_{groups}|}
\end{align*}
}
\normalsize
$G_{groups}$ is the 
set of all the groups that appear in the ground truth and $A_{groups}$ is the set for the approximate result.
This \textit{bin-completeness} (\S~\ref{sec:error-measure}) is an important measure for \texttt{GROUP BY} queries.
Fig.~\ref{fig:group-by} shows the bin-completeness numbers on top of the box plots. Higher is better. Both \name and DeepDB can consistently provide much higher bin-completeness, compared to other baselines. However, the error for \name is much better than DeepDB.

\subsection{Design Components Analysis}
\noindent\textbf{Samples Generated per Query.}
Generating more representative samples by \name can potentially improve accuracy, but at the cost of larger query latency and memory footprint. Table~\ref{tab:num-samples-variation} shows the change in the R.E. for \name as we increase the number of generated samples.
Generating 1000 representative samples, from the learned conditional distribution, is good enough to achieve a low AQP error, and beyond that the improvement is not significant.

\noindent\textbf{Impact of the Masking Strategy.}
Fig.~\ref{fig:masking_results} quantify the benefit of our proposed masking strategy for Flights dataset. We compare the proposed method with a \textit{No Masking} strategy where only \textit{NaN} input values, if any, are masked. We also compare with \textit{Random Masking} strategy that uses a uniform random masking, irrespective of the column type~\cite{vaeac}. We observe that the proposed stratified masking strategy can significantly reduce the AQP error compared to the other two alternative, irrespective of the number of predicates used in the queries.

\noindent\textbf{Masking factor.} Fig.~\ref{fig:masking-factor} shows the sensitivity of \name corresponding to the masking factor, across the range of predicates for the Flights dataset. A masking factor of 0.5 would be the choice, if we want to optimize for queries with arbitrary number of predicates.

\textit{We provide more results, latency and memory profiling numbers for \name along with discussion on future work in the supplement.} 
\section{Related Work}
\label{sec:rel_work}

\noindent\textbf{Approximate Query Processing}
has a rich history with the use of sampling and sketching based approaches ~\cite{AQP-not-silver-bullet, cormode2011sketch, chaudhuri2007optimized}.
BlinkDB~\cite{blinkdb} uses offline sampling, but assumes it has access to historical queries to reduce or optimize for storage overhead.
Related works used as baselines were described in \S~\ref{sec:eval}.

\noindent\textbf{Representative Data Generation.}
With the advancement of deep-learning, multiple generative methods have been proposed, primarily using GAN(s)~\cite{2019-gan-image-synthesis} and VAE(s)~\cite{vaeac}.
These techniques have been successful in producing realistic synthetic samples for images, auto-completing prompts for texts \cite{radford2019language, devlin2018bert}, generating audio wavelets \cite{oord2016wavenet}. Few works used GAN(s)~\cite{ctgan, tablegan} for tabular data generation either for providing data-privacy or to overcome imbalanced data problem. 

\noindent\textbf{Selectivity or Cardinality Estimation} includes several classical methods with synopsis structures (e.g., histograms, sketches and wavelets) to provide approximate answers~\cite{synopses_survey}. 
Recently, learned cardinality estimation methods have become common including using Deep Autoregressive models to learn density estimates of a joint data distribution ~\cite{naru,dl-selectivity-estimation-sigmod-2020,neurocard}, embedding and representation learning based cost and cardinality estimation ~\cite{sun2019end}. 

\section{Conclusion}
\label{sec:conclusion}
We presented a deep neural network based approximate query processing (AQP) system that can answer queries using a predicate-aware generative model at client-side, without processing the original data. Our technique learns the conditional distribution of data and generates targeted samples based on the conditions specified in the query. The key contributions of the paper are the use of conditional generative models for AQP and the techniques to reduce approximation error for queries with large number of predicates.
\bibliography{acmart}

\clearpage

\begin{appendices}
\section{Query Handling}
\label{sec:query_handling}
\name, like other recently proposed ML-based AQP techniques~\cite{dbest, deepdb, vae} targets the most dominant query templates for exploratory data analytics. In Section ``Target Query Structure", we provided high-level characteristic of queries that \name supports. Now we provide details about how exactly are these different classes of queries handled.

\noindent\textbf{Conjunction predicates on categorical and discrete attributes:}
E.g.
{\small
\texttt{\textbf{SELECT} $\Psi$, \textbf{AGG}($A_{N_i}$) \textbf{FROM} $T$ \textbf{WHERE} $A_{C_i}=p$ \textbf{AND} $A_{C_j}=q$}
}

A compound predicate composed of conjunctions (ANDs) in \texttt{WHERE} clauses is handled by using the specified predicate conditions on the discrete or categorical attributes as input to \name's conditional sample generator (masking other attributes) and then the query is executed on the generated samples to calculate the final answer.

\noindent\textbf{Disjunction predicates on categorical and discrete attributes:}
E.g.
{\small
\texttt{\textbf{SELECT} $\Psi$, \textbf{AGG}($A_{N_i}$) \textbf{FROM} $T$ \textbf{WHERE} $A_{C_i}=p$ \textbf{OR} $A_{C_i}=q$}
}

We support disjunctions (ORs) in WHERE conditions by decomposing it to multiple individual queries with each component of OR. We compute the \texttt{AVG} and \texttt{COUNT} of each individual query, and then use the \texttt{COUNT} as the weights to compute the overall \texttt{AVG}. For \texttt{SUM}, we directly compute the \texttt{SUM} of the individual queries.

If the OR conditions are on different variables, we use the principle of inclusion and exclusion
to compute the required aggregate. For example, to compute $\pi$ =
$( A_{C_i}=p$ \textbf{OR} $A_{C_j}=q )$, we will need to evaluate the queries, (1) $A_{C_i}=p$, (2) $A_{C_j}=q$ and (3) $( A_{C_i}=p$ \textbf{AND} $A_{C_j}=q )$, and then combine the results from these to obtain the result for the original query.

Note, \name can handle any combinations of \texttt{AND} and \texttt{OR} conditions by converting it to 
Disjunctive Normal Form (i.e. converting it to a structure of \texttt{[[.. AND ..] OR [.. AND ..]]}) using Propositional Logic. 

An \texttt{IN} (or \texttt{ANY}) condition can be similarly broken-down into multiple different predicates combined with \texttt{OR} and handled in the same manner.

\noindent\textbf{Range predicates on continuous attributes: }
E.g.
{\small
\texttt{\textbf{SELECT} $\Psi$, \textbf{AGG}($A_{N_i}$) \textbf{FROM} $T$ \textbf{WHERE} $A_{N_j}$ \textbf{BETWEEN} $p$ \textbf{AND} $q$}
}.
\name is designed primarily for conditions on discrete or categorical attributes, which is the dominant usecase for exploratory data analytics.
However, \name can handle range queries on continuous attributes by converting it into discrete categories through discretization (or binning) for training. During sample generation, the range condition is converted into multiple OR-ed conditions, one for each discrete categories, and samples are generated and merged as discussed before.
In the extreme case, \name falls back to condition-less generation or only use the partial conditions (i.e. using only the discrete or categorical predicates) to first generate a broader range of samples and then apply the range predicates, specified on the continuous attributes, on those samples to further filter the samples before calculating the aggregate value.

\noindent\textbf{Supporting GROUP BY: }
E.g.
{\small
\texttt{\textbf{SELECT} $\Psi$, \textbf{AGG}($A_{N_i}$) \textbf{FROM} $T$ \textbf{WHERE} $A_{C_i}=p$ \textbf{GROUP BY} $A_{C_j}$}
}

\name supports \texttt{GROUP BY} queries by decomposing it to multiple queries, one for each category of the categorical attribute on which the \texttt{GROUP BY} is done.

\noindent\textbf{Different aggregates --- AVG, SUM and COUNT: }
\name answers the queries with \texttt{AVG} as the aggregate function by directly computing average on the samples generated in a predicate-aware manner from its generative model. Similarly, using the conditional selectivity estimation model, it can directly answer \texttt{COUNT} queries using the predicate as the input. To answer \texttt{SUM} queries, \name first computes the average on the generated samples and then multiplies that with the selectivity estimate. 

\noindent\textbf{Supporting JOINs: }
\name can support JOINs for predictable/popular joined datasets
by first precomputing the JOIN-ed result table,
then building the generative and selectivity estimation models on this resulting table. This approach is similar to other ML-based AQP techniques~\cite{dbest, vae}.
 
\noindent\textbf{ORDER BY and LIMIT: }
Attributes that are used in \texttt{ORDER BY} can only be ones on which \texttt{GROUP BY} was applied or the attribute on which the aggregate was applied. Hence \name can handle these queries without treating them specially.
Low approximation error helps it mimic the ordering (and the set of records for \texttt{LIMIT}) as the fully accurate result. 

\section{Implementation Details}
\label{supp:implementation}
In this section, we provide more details about the workload, datasets and the training methodology.

\begin{figure}
\begin{minipage}{1.0\columnwidth}
  \begin{subfigure}{0.49\columnwidth}
    \centering
    \includegraphics[width=1.0\linewidth]{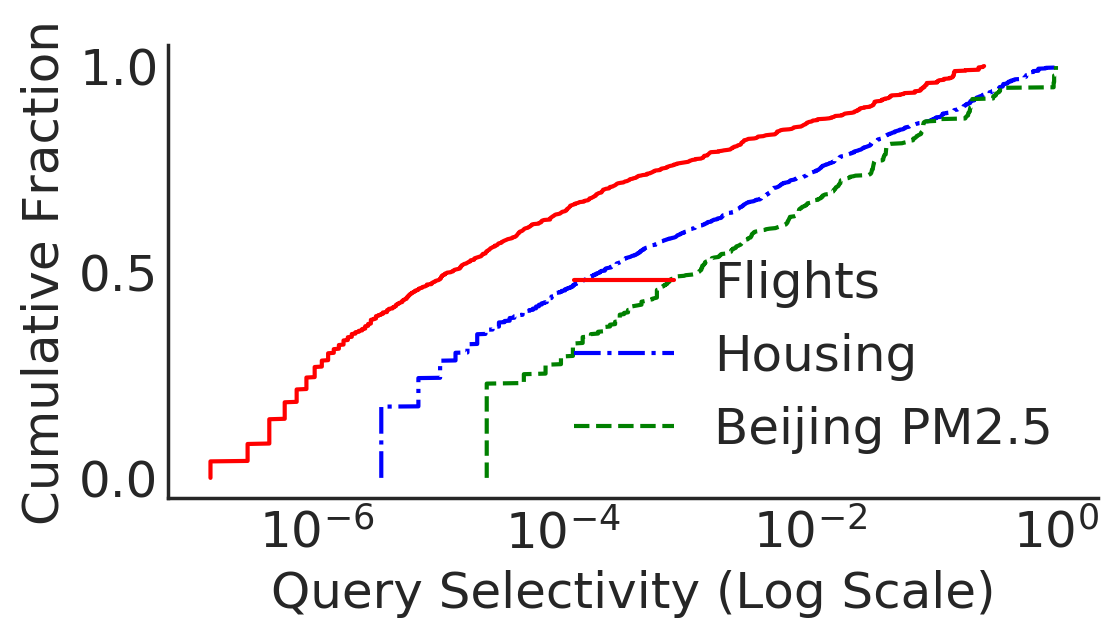}
    \caption{Synthetic queries}
    \label{fig:query_selectivity_synthetic}
  \end{subfigure}%
  \hfill
  \begin{subfigure}{0.49\columnwidth}
    \centering
    \includegraphics[width=1.0\linewidth]{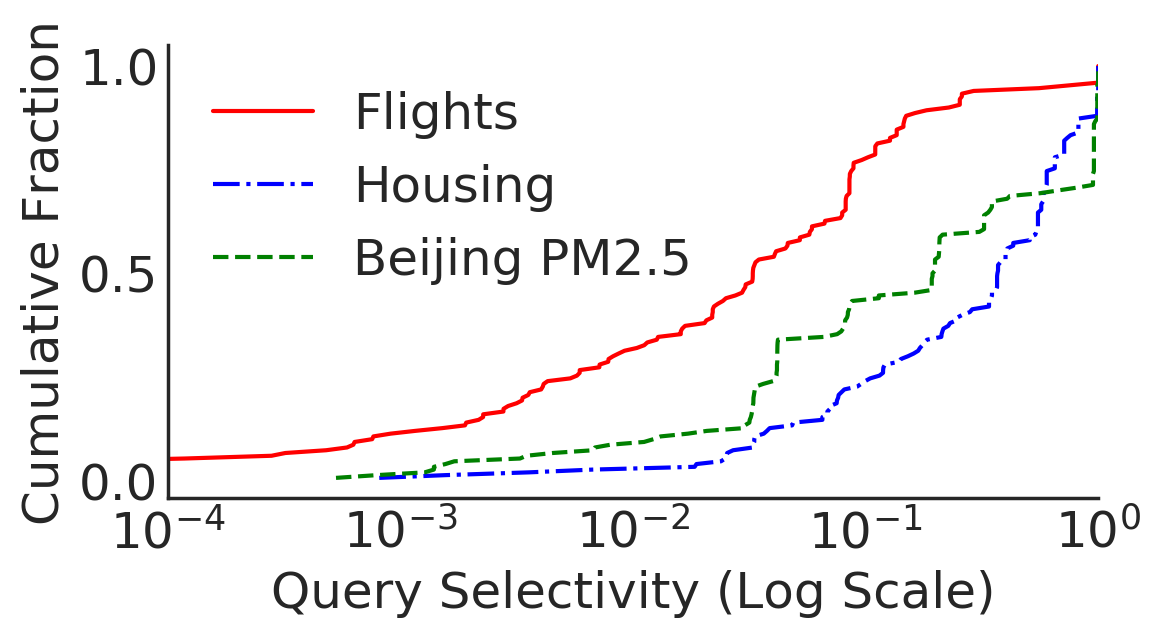}
    \caption{Production queries}
    \label{fig:query_selectivity_aep}
  \end{subfigure}
  \caption{Workload selectivity characteristics}
  \label{fig:query_selectivity}
\end{minipage}
\end{figure}

\begin{table}
\centering
\scalebox{0.8}{
\begin{tabular}{c|c|c|c}
\toprule
Dataset & \# of Rows & \# of Categorical & \# of Numerical\\
\midrule
Flights & 7.26M & 9 & 6\\
Housing & 299K & 14 & 4\\
Beijing PM2.5 & 43.8K & 7 & 5\\
\midrule
\end{tabular}
}
\caption{Details of Real-World Datasets}
\label{tab:datasets}
\end{table}

\subsection{Workload Characteristics}
As described in the Section ``Experimental Environment", the synthetic query workload had the following SQL query template:
\texttt{SELECT} \texttt{AVG}($n_{i}$) \texttt{FROM} T \texttt{WHERE} c1 = v1 \texttt{AND} ... \texttt{AND} ck = vk
Here $n_{i}$ is a numerical column, c1,...,ck are categorical columns and v1,...,vk are values taken by these categorical columns.

Fig. \ref{fig:query_selectivity} shows how both the synthetic workload and production workload cover a wide range of query selectivity on the three real-world datasets.

\subsection{Dataset Characteristics}
The datasets described in the Section ``Experimental Evaluation" had a mix of both categorical and numerical columns, ensuring that we train and evaluate \name on diverse attribute sets making sure we learn to predict the values (of numerical attributes) while conditioning on predicates (categorical attributes). Table \ref{tab:datasets} describes the dataset statistics.

\subsection{Training Methodology}
\label{sec:training}
Before training, we perform a set of pre-processing transformations on the dataset.
We label encode the categorical attributes to convert the distinct categories to integer labels which are then used to create the one-hot vector. The unique values associated with a categorical attribute are stored and used for two purposes - first, during inference to transform the input query predicates to their encoded labels for input to the model and second, to identify the different groups while answering a \texttt{GROUP BY} query on these categorical attributes. For the numerical attributes, with multi-modal distribution we perform mode specific normalization. We use Kernel Density Estimation (KDE) with Scott's~\cite{scott} bandwidth parameter selection to identify inflection points in the density estimates and then fit a Gaussian mixture. 
Numerical values are standard normalized using these obtained modes.

\section{Additional Experiments}

\begin{figure}
\centering
\begin{minipage}[t]{1.0\columnwidth}
\begin{minipage}[t]{0.495\columnwidth}
\centering
\includegraphics[width=1.0\textwidth]{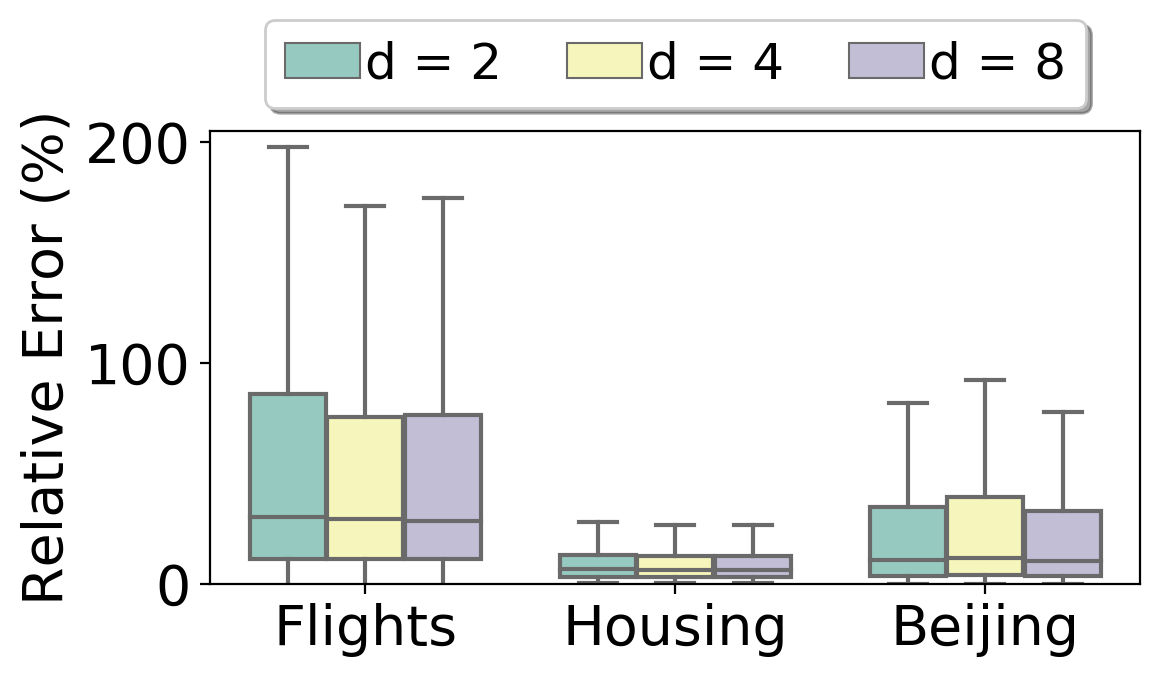}
\caption{Network Depths}
\label{fig:depth-variation}
\end{minipage}%
\hfill
\begin{minipage}[t]{0.495\columnwidth}
\includegraphics[width=1.0\textwidth]{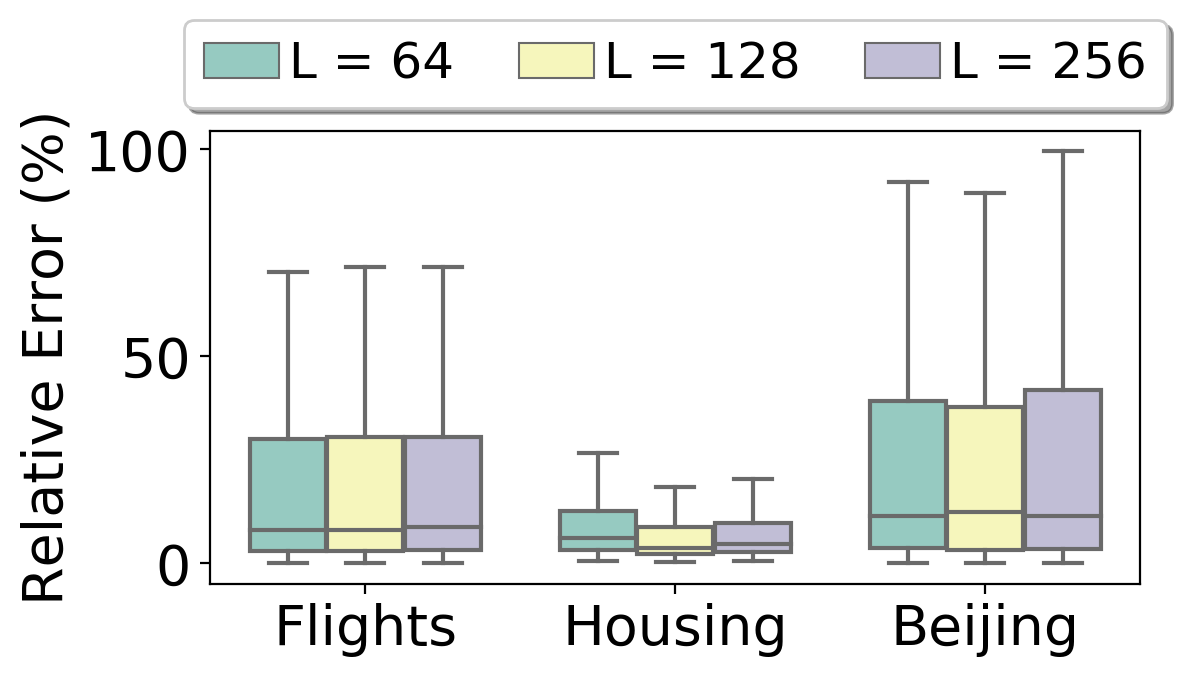}
\caption{Latent Dimensions}
\label{fig:ld-variation}
\end{minipage}
\end{minipage}
\end{figure}

\subsection{Depth of Prior \& Proposal Networks}
In Fig.~\ref{fig:depth-variation} we show the variations of AQP error, accumulated over all the queries with different number of predicates and for all three datasets. We set the depth ($d$) of both prior and proposal networks as 2, 4 and 8.
While more depth and hence more complex network helps to achieve better accuracy, it can be observed it has a diminishing returns. More depth also increases the size of the model and requires longer time for the training to converge. 

\subsection{Latent Dimension}
Latent dimension ($L$) was varied in Fig.~\ref{fig:ld-variation} between 64, 128 and 256. While in theory a larger latent dimension would help to capture the conditional distribution at a finer resolution, we observed that 128 is the optimum value across all datasets.

\subsection{Model Scalability}

\begin{figure}
    \centering
    \includegraphics[width=0.75\linewidth]{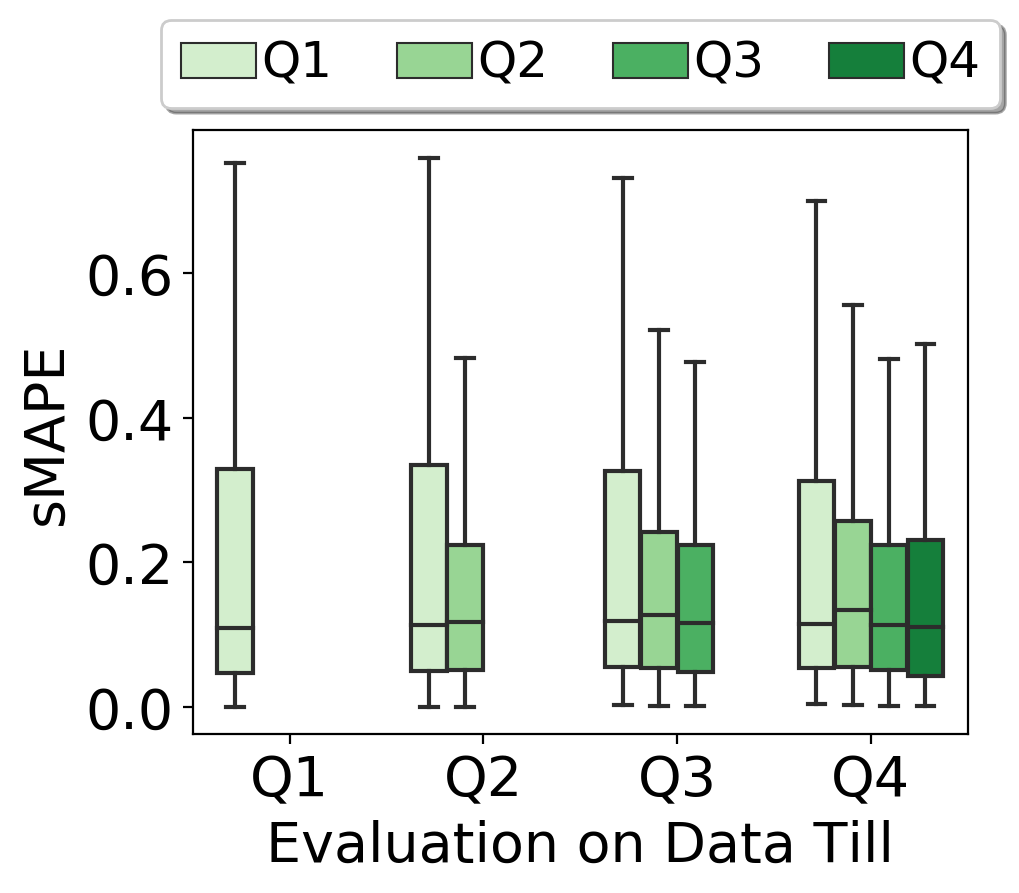}
    \caption{Performance of \name with increase of data with new quarterly data ingestion for Flights dataset. We show how model trained on Q1 performs on data for Q1+Q2, Q1+Q2+Q3 and Q1+Q2+Q3+Q4, how model trained on Q1+Q2 performs on Q1+Q2+Q3 and Q1+Q2+Q3+Q4 and so on. \textit{Same model-size} can keep the median error low with retraining, even when data size increases by 4 times.}
    \label{fig:w_wo_retraining}
\end{figure}

\begin{table}[ht]
    \centering
    \scalebox{0.8}{
        \begin{tabular}[t]{>{\itshape} l|c|c|c|c}
            \toprule
            \thead{Data} & \thead{\# Rows} & \thead{mins/epoch} & \thead{Model size}\\
            \midrule
            Q1 & 1.7M & 3.19 & 15.08 MB\\
            Q1-Q2 & 3.5M & 6.18 & 15.08 MB\\
            Q1-Q3 & 5.4M & 8.56 & 15.08 MB\\
            Q1-Q4 & 7.2M & 11.41 & 15.08 MB\\
            \midrule
        \end{tabular}
    }
    \caption{Changes of data-size, re-training time and model-size.}
    \label{tab:scalability-variation}
\end{table}

\begin{figure*}[t]
    \centering
    \begin{minipage}[t]{1.0\textwidth}
    \centering
    \vspace{0pt}
    \begin{subfigure}[t]{0.33\textwidth}
      \includegraphics[width=1.0\linewidth]{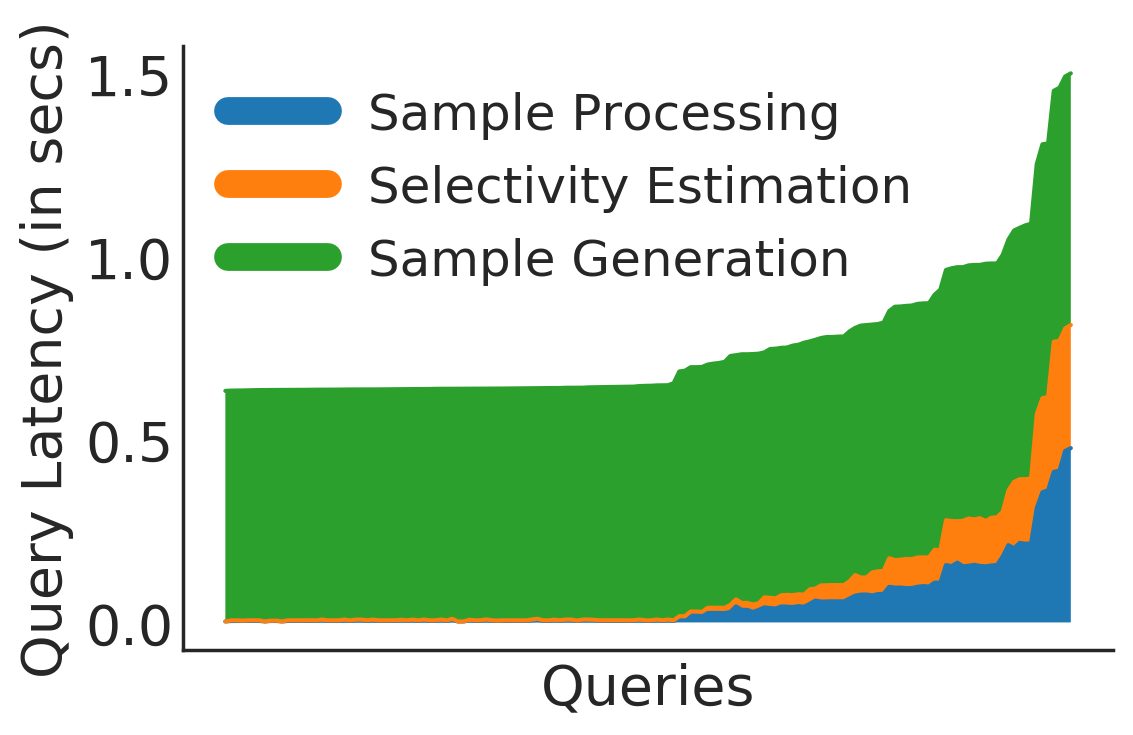}
      \caption{Flights}
      \label{fig:query-latency-flights}
    \end{subfigure}%
    \hfill
    \begin{subfigure}[t]{0.33\textwidth}
      \includegraphics[width=1.0\linewidth]{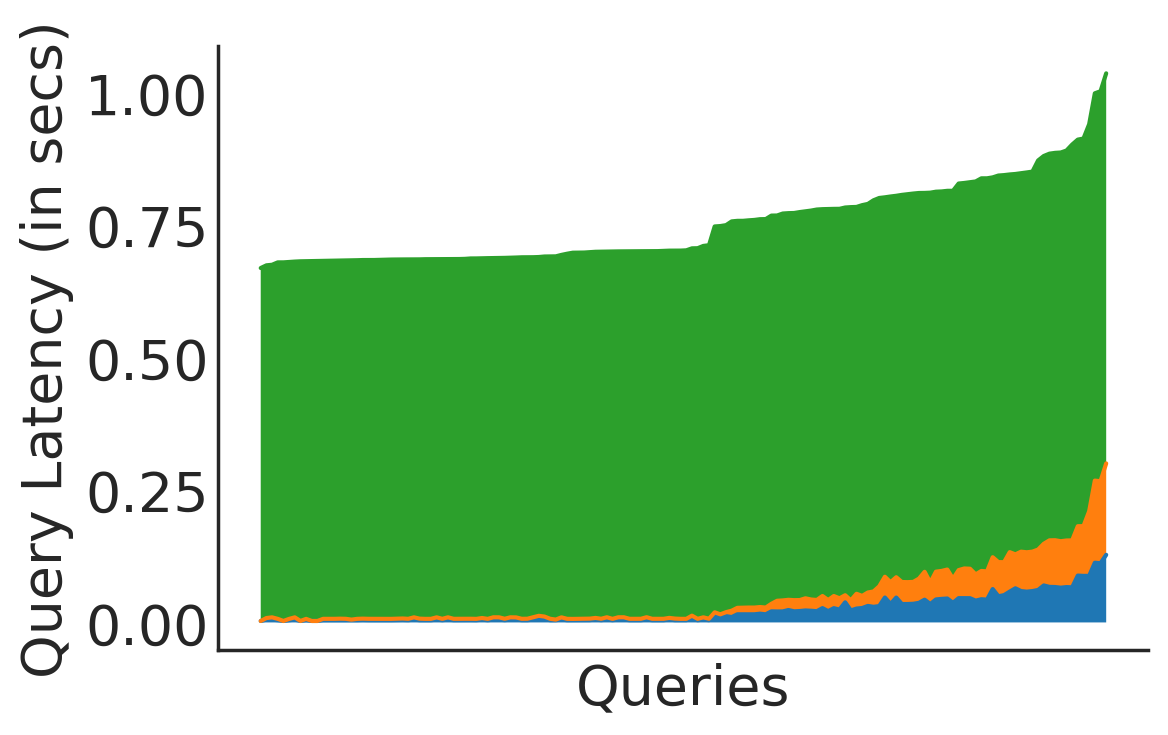}
      \caption{Housing}
      \label{fig:query-latency-housing}
    \end{subfigure}%
    \hfill
    \begin{subfigure}[t]{0.33\textwidth}
      \includegraphics[width=1.0\linewidth]{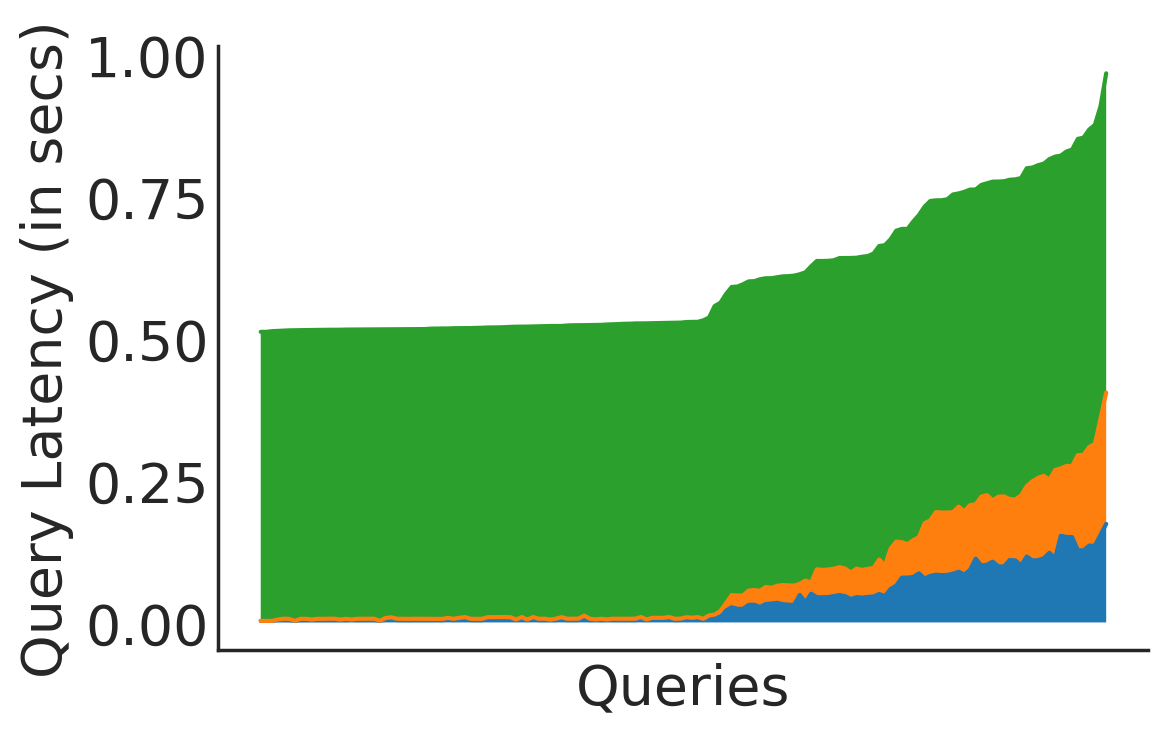}
      \caption{Beijing PM 2.5}
      \label{fig:query-latency-pm2.5}
    \end{subfigure}%
    \caption{Query latency on production workload}
    \label{fig:query-latency}
    \end{minipage}
\end{figure*}

In many real-world scenario, newly available data is periodically ingested to the system in batch and appended to the existing data on which analysts run queries to explore and find insights. 
In this section, we evaluate how the AQP error for \name changes if we use the existing models to answer the queries that are targeted towards the whole of the newly appended data.

We do this using the Flights dataset and divide it into four quarters Q1-Q4 of 2019, using the already available \textit{Quarter} column. Then we train \name only with data till $Q_{i}$ where $i \in [1,4]$ and then evaluate using ground truth till $Q_{j}$ where $j \in [i,4]$.
As shown in Fig.~\ref{fig:w_wo_retraining}, 
we found \name's old models were able to keep the median AQP error low (with only some increase in the outliers), but the tail of the error-distributions increased when evaluated on the new data that is being progressively appended. 
Please note, in Fig.~\ref{fig:w_wo_retraining}, the {$Q_{i}$, $Q_{i}$} bars show that how much the AQP errors would reduce if \name performs a \textit{retraining} on the newly appended data. 
We can also see from the median latency plots, that retraining need not be performed for every data ingestion as the previously trained model can sustain low approximation error up to a mark and gradually increasing for additional fresh data ingestion. 
Needless to say, this observation is very much dependent on the characteristics of the actual data sources and \name's design has a module to specifically detect such shifts in the data distribution as part of its ETL logic for (re-)training.

In Table \ref{tab:scalability-variation}, we also show the variation in training time per epoch for different dataset sizes due to data ingestion. The models were trained for 50 epochs within which the VLB loss had converged. A similar convergence behavior was observed on other datasets. Note that the same model-size (15.08 MB) can bring down the error after retraining on data upto 4x large, while the training time per epoch linearly increases with the training data size. 
The model-size did not increase because hyperparameters of the architecture as well as the number of categories in the dataset did not change.
In order to tame this for larger datasets, one can train on a sample of the original data depending on the time and resource availability.

Scalability of \name model training phase can also be improved using parallel or distributed learning supported by all major deep-learning frameworks like Tensorflow and PyTorch.  
Other general distributed learning algorithms e.g.,  
DownpourSGD, ADMM, EASGD, and
GoSGD~\footnote{Well summarized by Zhang in ``Distributed stochastic optimization for deep learning", PhD thesis New York University, 2016}
can also be used to extend our approach for handling very large datasets.  
\name can be trained even on a sampled data without much loss of accuracy.
Finally, in case there are any attributes that are very frequently used, then original data can also be divided based on different categories of that attribute and multiple models, one corresponding to each part can be trained in parallel.

\subsection{Query Latency and Memory Footprint}
In this section we characterize \name's query processing latency, how different submodules contribute to that latency, along with its runtime memory footprint. 

\noindent\textbf{Latency: }
Figure~\ref{fig:query-latency-flights}, ~\ref{fig:query-latency-housing},~\ref{fig:query-latency-pm2.5} shows the end-to-end query latency for the production workload queries (sorted according to latency), also the time taken by different parts.
Recall, \name's runtime workflow has 3 components: sample generation, sample processing and selectivity estimation.
Sample processing involves calculating an average by executing the query on the initial 1000 samples generated.
As discussed in \S~\ref{sec:query_handling}, \name decomposes a query into multiple subqueries for \texttt{GROUP BY}s and for \texttt{OR}, \texttt{IN} operators in \texttt{WHERE} conditions.
Selectivity estimation is needed for both handling
\texttt{SUM} as the aggregate, as well as for combining the results from multiple subqueries with appropriate weights.
The queries towards the end are queries with large number of sub-queries increase, leading to higher latency. Particularly, true for Flights data where \texttt{GROUP BY} queries resulted in a larger number of subqueries due to columns with 50+ or even 360+ unique categories.

\begin{table}[ht]
\centering
\scalebox{0.8}{
\begin{tabular}[t]{l|c|c|c}
\toprule
   Dataset & Selectivity Estm. & Sample Gen. & Sample Processing\\
   \midrule
   Flights & 21.2 & 16.25 & 6.9 \\
   Housing & 11.23 & 10.25 & 4.85 \\
   Beijing PM2.5 & 10.78 & 10.48 & 4.43 \\
   \midrule
   \end{tabular}
}
\caption{Runtime Memory Usage (All numbers in MB).}
\label{tab:memory-consumption}
\end{table}

\noindent\textbf{Reducing Query Latency.}
\name's model size as well as inference latency can further be improved using post-training quantization techniques~\cite{quantization} available from deep-learning frameworks.
We found that converting \name's model to ONNX format~\cite{onnx} (for running with Javascript, within a web-browser) can also significantly reduce its inference latency.
Multiple sub-queries corresponding to disjunctions or \texttt{GROUP BY}s can potentially also be handled in parallel by different cores at the client-side to further reduce latency. 
A caching layer can be introduced at the client-side, to cache samples generated for frequently used combination of predicates latency improvements.

\noindent\textbf{Memory Consumption.}
Peak runtime memory consumption was measured during the 3 stages of operation for \name. using \texttt{@profile} decorator from python \texttt{memory profiler} \cite{memprof} module.
From Table~\ref{tab:memory-consumption} it can be observed that peak memory was about 21.2 MB for Flights, 11.2 MB for Housing and 10.8 MB for Beijing PM 2.5. Across all datasets and memory peaks during selectivity estimation (S. E.) using \naru.
Memory usage of Flights is higher because of some categories had large number of unique values that required higher one hot encoding dimension by \name. 
Please note, that \name can keep low runtime memory footprint, as it only generates a small number of sample on the fly to answer the queries and does not keep a large number of samples in memory.

\noindent\textbf{Discussion: }
Please note, achieving the \textit{lowest latency} is not our goal, rather lowering the approximation error for complex but critical queries is our goal, as long as the latency is still within the bounds of \textit{humans-speed} for good user-experience.
After talking to several experienced analysts, we understand $<1$ second latency, which we achieve for most of the queries (Fig.~\ref{fig:query-latency}), is well within the bounds of human-speed for insight discovery use-cases. Having said that, we see significant scopes for improving latency by converting to special model formats (e.g. ONNX~\cite{onnxjs}), using quantization~\cite{quantization} and parallelism across cores --- including parallel inference between our CVAE and \naru models. None of these we have used in this paper. Quantization, pruning and other neural network compression techniques~\cite{deep-compression} can also significantly reduce the model-size and hence the run-rime memory footprint. We leave it as a future work to characterize the approximation errors, latency and memory footprint for all ML-based AQP techniques, including \name, when neural-network compression techniques and other forms of regularization are applied. 
\section{Discussion and Future Work}
\label{sec:discussion}

We now discuss limitations of \name and our future works.

\noindent\textbf{Confidence Intervals (CI): }
We plan to focus on developing a CI calculation method for \name. While providing an accurate CI is challenging for all AQP systems~\cite{AQP-not-silver-bullet}.
Some AQP systems~\cite{blinkdb, deepdb} provide a CI assuming a normal distribution for the predicted results. \name might use a similar approach.
But, we also plan to explore how bootstrapping~\cite{diciccio1996bootstrap} can be used to repeatedly sample the latent-space of our CVAE model to calculate a CI, which has never been done for a DNN-based generative model.  

\noindent\textbf{Additional Aggregates: }
\name now supports \texttt{AVG}, \texttt{SUM}, \texttt{COUNT} aggregates over numerical columns.
In the future, we plan to extend our technique to support to more type of aggregates such as \texttt{MAX}, \texttt{MIN} and \texttt{COUNT DISTINCT}.
For \texttt{COUNT DISTINCT}, some promising approximation techniques are available using a combination of CountMin and HyperLogLog sketches~\cite{ting2019approximate}. We will investigate, how such techniques can be used at the client-side and integrated with \name alongside or replacing \naru for cardinality estimation.
Capturing MAX and MIN in a conditional generative model is a non-trivial problem. We will explore how to incorporate the concepts of extreme quantile estimation~\cite{beirlant2004nonparametric} in the training process of our conditional generative model. 

\noindent\textbf{Incremental Retraining: }
As observed in Table~\ref{tab:scalability-variation}, the retraining time of \name scales linearly with size of data after fresh ingestion data batches.
We observed that for very large datasets, \name trained only on a small fraction of uniform samples (5-10\%) can still answer queries with comparable relative error as the original data. If hints about important attributes in the queries are known a priori, such training can also be performed using a stratified sample.
Such training over samples can significantly reduce the retraining-time when data-size keeps on growing. 
However, we plan to investigate how \name's model can be incrementally trained on the new data without \textit{catastrophic forgetting}~\cite{lee2017overcoming}.  
Transfer learning~\cite{zhuang2020comprehensive} for generative or conditional generative models is also an untapped area, worth pursuing in this context.

\noindent\textbf{No universal winner --- Hybrid AQP System: }
\name's system design, as described in Section ``Design of \name" allows the use of a server-side exact query processing mechanism for the query structures that \name currently does not support (Ref. \S~\ref{sec:query_handling}).
However, a more intelligent design of a client-side hybrid AQP system should be investigated as well.  
The reason being, as observed from Figure 4 (AQP Error vs \# of predicates), in the spectrum of queries with 1 to 9 (or more predicates), there is no universal winner. 
While DeepDB is the best-performing client-side technique for queries with 1-2 predicates, \name is the best-performing technique for queries with more than 3 predicates. 
Similar to recently proposed mobile computer vision systems that keep multiple object classifiers or detectors to optimize for accuracy and latency for different video frames~\cite{han2016mcdnn}, we can also explore how a hybrid ML-based AQP system can use different AQP models, depending on the query properties to optimize for accuracy and latency.
\end{appendices}

\end{document}